\documentclass{article}
\usepackage{fullpage}
\usepackage{amssymb,amsthm,amsmath}
\usepackage{authblk}
\usepackage[utf8]{inputenc}
\usepackage[noend,boxed, noline]{algorithm2e}
\SetKwComment{Comment}{$\triangleright$\ \rm}{}
\usepackage[hidelinks]{hyperref}
\usepackage[sort,capitalize]{cleveref}
\usepackage{enumerate}
\usepackage{thmtools}
\usepackage{thm-restate}
\usepackage{microtype}
\usepackage{todonotes}
\usepackage{comment}
\usepackage{color} 

\usepackage{tikz}
\usetikzlibrary{shapes,decorations.pathreplacing,calc}

 \newif\ifdraft\drafttrue
\def\timestamp{%
    \the\year-%
    \ifnum\month<10 0\fi\the\month-%
    \ifnum\day<10 0\fi\the\day\ {%
      \count9=\time \divide\count9 by 60 %
      \ifnum\count9<10 0\fi\the\count9%
      \count7=\time \multiply\count9 by60 \advance\count7 by-\count9%
      :%
      \ifnum\count7<10 0\fi\the\count7%
    }
}
 \ifdraft
\usepackage{fancyheadings}
\fi

\newtheorem{theorem}{Theorem}[section]
\newtheorem{observation}[theorem]{Observation}
\crefname{observation}{Observation}{Observations}
\newtheorem*{claim}{Claim}
\newtheorem{fact}[theorem]{Fact}
\newtheorem{corollary}[theorem]{Corollary}
\newtheorem{definition}[theorem]{Definition}
\newtheorem{lemma}[theorem]{Lemma}
\theoremstyle{remark}
\newtheorem{remark}[theorem]{Remark}
\newtheorem{example}[theorem]{Example}

\newcommand{\Oh}{\mathcal{O}}
\newcommand{\ceil}[1]{\lceil #1 \rceil}

\newcommand{\al}{\mathtt{a}}
\newcommand{\bl}{\mathtt{b}}

\renewcommand{\L}{\mathcal{L}}

\newcommand{\sub}{\subseteq}

\newcommand{\Occ}{\mbox{\textit{Occ}}}

\newcommand{\Compress}{\mathsf{COMPR}}

\newcommand{\Fa}{\mathcal{F}}
\newcommand{\refine}{\mathsf{refine}}

\newcommand{\F}{\mathsf{SUB}}

\newcommand{\Seeds}{\mathsf{Seeds}}

\newcommand{\FSeeds}{\textrm{SEEDS}\xspace}
\newcommand{\LSeedsRep}{\textrm{LONG-SEEDS}\xspace}
\newcommand{\BSeedsRep}{\textrm{INT-SEEDS}\xspace}
\newcommand{\Combine}{\textrm{Combine}}
\newcommand{\Paths}{\textrm{PATHS}\xspace}

\newcommand{\findop}{\mathtt{find}}
\newcommand{\unionop}{\mathtt{union}}
\renewcommand{\SS}{\mathcal{S}}

\newcommand{\pack}{\mathsf{pack}}
\newcommand{\PACK}{\mathsf{PACK}}
 \newcommand{\LCands}{\mathsf{LCands}}
 \newcommand{\RCands}{\mathsf{RCands}}
 \newcommand{\Seedseeds}{\mathsf{QSeeds}}
\DeclareMathOperator*{\id}{id}

 \newcommand{\maxgap}{\mathsf{maxgap}}

\newcommand{\AAlph}{\mathit{Alph}}

\def\dotdot{\mathinner{\ldotp\ldotp}}

  \author[1,2]{Tomasz Kociumaka}
  \author[1]{Marcin Kubica}
  \author[1]{Jakub Radoszewski}
  \author[1]{Wojciech Rytter}
  \author[1]{Tomasz~Wale\'n}

\affil[1]{\normalsize Institute of Informatics,
University of Warsaw, Poland\\
\texttt{[kociumaka,kubica,jrad,rytter,walen]@mimuw.edu.pl}}
\affil[2]{\normalsize Department of Computer Science, Bar-Ilan University, Ramat Gan, Israel}
\date{\vspace{-1cm}}

\begin{document}

\title{A Linear Time Algorithm for Seeds Computation}

\maketitle
\begin{abstract}
  A seed in a word is a relaxed version of a period in which the occurrences of the repeating subword may overlap.
  We show a linear-time algorithm computing a linear-size
  representation of all the seeds of a word (the number of seeds might be quadratic). 
  In particular, one can easily derive the shortest seed and the number of seeds from our representation.
  Thus, we solve an open problem stated in the survey by Smyth~(2000) and improve
  upon a previous $\Oh(n\log n)$ algorithm by Iliopoulos, Moore, and Park~(1996).
  Our approach is based on
  combinatorial relations between seeds and subword complexity (used here for
  the first time in context of seeds).
  In the previous papers, the compact representation of seeds consisted of two independent
  parts operating on the suffix tree of the word and the suffix tree of the reverse of the word, respectively.
  Our second contribution is a simpler representation of all seeds which avoids dealing with the reversed word.

  A preliminary version of this work, with a much more complex algorithm constructing the earlier representation of seeds, 
   was presented at the 23rd Annual ACM-SIAM Symposium of Discrete Algorithms (SODA 2012).
\end{abstract}

\section{Introduction}
  The notion of periodicity in words is widely used in many fields, such as
  combinatorics on words, pattern matching, data compression, automata theory,
  formal language theory, molecular biology etc.\ (see~\cite{Lothaire2005}).
  The concept of quasiperiodicity, introduced by Apostolico and Ehrenfeucht in~\cite{DBLP:journals/tcs/ApostolicoE93},
  is a generalization of the notion
  of periodicity:  A quasiperiodic word is entirely covered by occurrences of another
  (shorter) word, called the \emph{quasiperiod} or the \emph{cover}.
  The occurrences of the quasiperiod may overlap, while in a periodic repetition
  the occurrences of the period do not overlap.
  Hence, quasiperiodicity enables detecting repetitive structure of words
  which cannot be found using the classic characterizations in terms of periods.
  
  \begin{figure}[b!]
      \begin{center}
\begin{tikzpicture}[scale=0.6]
\clip (-0.15,-0.41) rectangle (8.65,0.96);
\foreach \x in {0,1,3,4,6,8,9,11,13,14,16,17} {
\draw (\x/2,0) node [above] {{$\al$}};
}
\foreach \x in {2,5,7,10,12,15} {
\draw (\x/2,0) node [above] {{$\bl$}};
}
\
\foreach \x in {-2,4,9,14} {
   \draw[thick,yshift=0.25cm] (\x/2-0.2,0.5) -- (\x/2-0.2,0.6) -- (\x/2+1.2,0.6) -- (\x/2+1.2,0.5);
}
\foreach \x in {1,6,11,17} {
   \draw[thick] (\x/2-0.2,0) -- (\x/2-0.2,-0.1) -- (\x/2+1.2,-0.1) -- (\x/2+1.2,0);
}
\end{tikzpicture}
  \end{center}
  \begin{center}
    \begin{tikzpicture}[scale=0.6]
\clip (-0.125,-0.41) rectangle (8.65,0.96);
\foreach \x in {0,1,3,4,6,8,9,11,13,14,16,17} {
\draw (\x/2,0) node [above] {{$\al$}};
}
\foreach \x in {2,5,7,10,12,15} {
\draw (\x/2,0) node [above] {{$\bl$}};
}
\
\foreach \x in {-2,6,14} {
   \draw[thick,yshift=0.15cm] (\x/2-0.2,0.5) -- (\x/2-0.2,0.6) -- (\x/2+2.2,0.6) -- (\x/2+2.2,0.5);
}
\foreach \x in {1,11} {
   \draw[thick] (\x/2-0.2,0) -- (\x/2-0.2,-0.1) -- (\x/2+2.2,-0.1) -- (\x/2+2.2,0);
}
\end{tikzpicture}
  \end{center}
    \vspace*{-0.7cm}
    \caption{\label{fig:ex_seed}
      The word $\texttt{aba}$ (above) is the shortest seed of the word $w=\texttt{aabaababaababaabaa}$.
      Another seed of $w$ is $\texttt{abaab}$ (below). Two of its ``overhanging'' 
occurrences correspond to boundary subwords $\texttt{aab}$ and $\texttt{abaa}$.
 In total, the word $w$ has 35 distinct seeds, but does not have a non-trivial cover.
    }
  \end{figure}

  An extension of the notion of a cover is the notion of a \emph{seed}: a cover which is
  not necessarily aligned with the ends of the word being covered, but is allowed
  to overflow on either side; see \cref{fig:ex_seed}.
  More formally,
  a word $v$ is a \emph{seed} of $w$ if $v$ is a subword of $w$ and $w$ is a subword
  of some word $u$ covered by $v$.

  Seeds were first introduced and studied by Iliopoulos, Moore, and Park~\cite{DBLP:journals/algorithmica/IliopoulosMP96}.
  The original motivation for covers and seeds comes from DNA sequence analysis (the search for regularities and common features
  in DNA sequences).
  Due to natural applications in molecular biology (a hybridization approach to
  analysis of a DNA sequence), both covers and seeds have also been extended in the sense
  that a number of factors are considered instead of a single word \cite{IliopoulosSmythKCovers}.
  This way, the notions of
  $k$-covers \cite{DBLP:journals/jalc/ColeIMSY05,DBLP:journals/isci/IliopoulosMS11},
  $\lambda$-covers \cite{DBLP:journals/isci/GuoZI07}, and
  $\lambda$-seeds \cite{DBLP:conf/aaim/GuoZI06} were introduced.
  In applications such as molecular biology and computer-assisted music
  analysis, finding exact repetitions is not always sufficient; the same problem
  occurs for quasiperiodic repetitions.
  This led to the introduction of the notions of approximate covers and seeds \cite{DBLP:conf/cpm/AmirLLLP17,DBLP:conf/cpm/AmirLLP17,DBLP:conf/cpm/AmirLP18,SimParkKimLee,DBLP:journals/jalc/ChristodoulakisIPS05}, partial covers and seeds \cite{DBLP:journals/tcs/FlouriIKPPST13,DBLP:journals/algorithmica/KociumakaPRRW15,DBLP:journals/tcs/KociumakaPRRW16}, and approximate $\lambda$-covers~\cite{DBLP:conf/spire/GuoZI06}.

  \paragraph{Previous results.}
  Iliopoulos, Moore, and Park \cite{DBLP:journals/algorithmica/IliopoulosMP96}
  gave an $\Oh(n\log{n})$-time algorithm computing a linear representation of all the seeds of a given word.
  For the next 15 years, no $o(n\log{n})$-time algorithm was known for this problem.
  Smyth formulated computing all the seeds of a word in linear time as an open problem
  in his survey~\cite{DBLP:journals/tcs/Smyth00}.
  Berkman et al.~\cite{DBLP:journals/iandc/BerkmanIP95} gave a parallel algorithm computing all the seeds in $\Oh(\log{n})$ time and
  $\Oh(n^{1+\varepsilon})$ space (for any positive $\varepsilon$) using $n$ processors in the CRCW PRAM model.
  Much later, Christou et al. \cite{DBLP:journals/tcs/ChristouCIKPRRSW13} proposed an alternative sequential $\Oh(n\log{n})$-time algorithm for computing the shortest seed.

  In contrast, a linear-time algorithm finding the shortest cover of a word was given
  by Apostolico et al.\ \cite{DBLP:journals/ipl/ApostolicoFI91} and later on improved
  into an on-line algorithm by Breslauer \cite{DBLP:journals/ipl/Breslauer92}.
  Moore and Smyth \cite{DBLP:journals/ipl/MooreS94,DBLP:journals/ipl/MooreS95} proposed a linear-time algorithm computing all the covers of a word,
  whereas Li and Smyth \cite{DBLP:journals/algorithmica/LiS02} afterwards developed an on-line algorithm for the problem of representing all covers of all prefixes of a word.

  Another line of research is finding maximal quasiperiodic subwords of a word.
  This notion resembles maximal repetitions (runs) in a word
  \cite{DBLP:conf/focs/KolpakovK99}, which is another widely studied notion of combinatorics
  on words.
  Two $\Oh(n\log{n})$-time algorithms for reporting all maximal quasiperiodic subwords
  of a word of length $n$ have been proposed by Brodal and Pedersen
  \cite{DBLP:conf/cpm/BrodalP00} and Iliopoulos and Mouchard
  \cite{DBLP:journals/jalc/IliopoulosM99}; these results improved upon the initial
  $\Oh(n\log^2{n})$-time algorithm by Apostolico and Ehrenfeucht~\cite{DBLP:journals/tcs/ApostolicoE93}.
  
  \paragraph{Our result.}
  We present a linear-time algorithm computing the set $\Seeds(w)$ of all seeds of a given word $w$.
  As illustrated in~\cref{ex:quadratic}, the number of seeds can be quadratic in the length $|w|$
  (contrary to the number of covers, which is always linear).
  Consequently, our algorithm returns a linear-size \emph{package representation} of the set $\Seeds(w)$,
  which allows finding a shortest seed and the number of all seeds in a very simple way.

  \begin{example}\label{ex:quadratic}
    The following word of length $4m+3$ contains $\Theta(m^2)$ different seeds:
    \[w=\al^m \bl \al^m \bl \al^m \bl \al^m.\]
    Those seeds are $\al^i \bl \al^j$ with $i+j\ge m$ and $0\le i,j\le m$.
  \end{example}

  Our procedure assumes that the alphabet $\Sigma$ of the input word $w$ consists of integers
  that are polynomial in terms of the length $n$ of $w$.

  \paragraph{Package Representation of Seeds.}
  For a word $w$, by $w[i \dotdot j]$ we denote the subword $w[i] \cdots w[j]$.
  We introduce packages which are collections of 
consecutive prefixes of a subword of $w$.
  More formally, for positive integers $i \le j_1\le j_2$, we define a \emph{package}:
  \[\pack(i,j_1,j_2)=\{w[i\dotdot j] : j_1\le j \le j_2\}.\]
  If $\L$ is a set of triples of integers, then we denote:
  \[\PACK(\L)  = \bigcup_{(i,j_1,j_2) \in \L} \pack(i,j_1,j_2).\]
  The output of our algorithm, called the \emph{package representation} of the set $\Seeds(w)$, consists of a 
  set $\L$ of integer triples such that
  $\Seeds(w) = \PACK(\L)$
  and all the packages in the representation are pairwise disjoint. (In other words, each seed belongs to exactly one package).

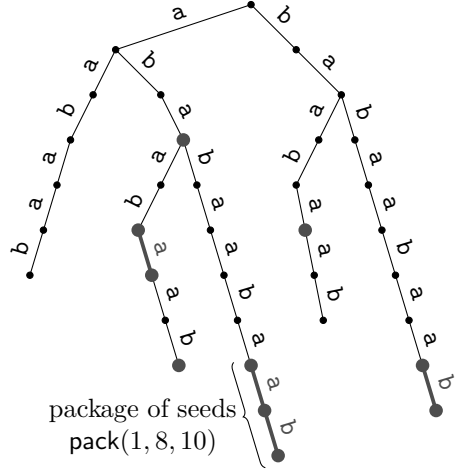
\begin{figure}[t]
  \begin{center}
    \newcommand{\edgeA}[5] {
  \draw (#1,#2) -- node[above, sloped] {$#5$} (#3,#4);
  \filldraw (#3,#4) circle (0.07cm);
}

\newcommand{\edgeAt}[5] {
  \draw[line width=0.05cm,white!30!black] (#1,#2) -- node[above,sloped] {$#5$} (#3,#4);
  \filldraw (#3,#4) circle (0.07cm);
}

\begin{tikzpicture}[scale=0.6]
  \filldraw (0,0) circle (0.07cm);
  \edgeA{0}{0}{-3}{-1}{\al}
  \edgeA{0}{0}{1}{-1}{\bl}

  \edgeA{-3}{-1}{-3.5}{-2}{\al}
  \edgeA{-3.5}{-2}{-4}{-3}{\bl}
  
  
  \edgeA{-4}{-3}{-4.3}{-4}{\al}
  \edgeA{-4.3}{-4}{-4.6}{-5}{\al}
  \edgeA{-4.6}{-5}{-4.9}{-6}{\bl}

  \edgeA{-3}{-1}{-2}{-2}{\bl}
  
  \edgeA{-2}{-2}{-1.5}{-3}{\al}

  \edgeA{-1.5}{-3}{-1.2}{-4}{\bl}
  \edgeA{-1.2}{-4}{-0.9}{-5}{\al}
  \edgeA{-0.9}{-5}{-0.6}{-6}{\al}
  \edgeA{-0.6}{-6}{-0.3}{-7}{\bl}
  \edgeA{-0.3}{-7}{0}{-8}{\al}

  \edgeA{-1.5}{-3}{-2}{-4}{\al}
  \edgeA{-2}{-4}{-2.5}{-5}{\bl}

  \edgeA{-2.2}{-6}{-1.9}{-7}{\al}
  \edgeA{-1.9}{-7}{-1.6}{-8}{\bl}


  \edgeA{1}{-1}{2}{-2}{\al}
  

  \edgeA{2}{-2}{1.5}{-3}{\al}
  \edgeA{1.5}{-3}{1}{-4}{\bl}
  

  \edgeA{1}{-4}{1.2}{-5}{\al}
  \edgeA{1.2}{-5}{1.4}{-6}{\al}
  \edgeA{1.4}{-6}{1.6}{-7}{\bl}
  
  \edgeA{2}{-2}{2.3}{-3}{\bl}
  \edgeA{2.3}{-3}{2.6}{-4}{\al}
  \edgeA{2.6}{-4}{2.9}{-5}{\al}
  \edgeA{2.9}{-5}{3.2}{-6}{\bl}
  \edgeA{3.2}{-6}{3.5}{-7}{\al}
  \edgeA{3.5}{-7}{3.8}{-8}{\al}

  \edgeAt{3.8}{-8}{4.1}{-9}{\bl}


  \fill[white!30!black] (3.8,-8) circle (0.15cm);
  \fill[white!30!black] (4.1,-9) circle (0.15cm);

  \draw[decorate, decoration=brace,xshift=-0.333cm,yshift=-0.1cm] (0.65,-10.1666) -- (-0.05,-7.8333);

  \draw[xshift=-0.4cm] (-2,-9) node (A) {package of seeds};
  \draw[xshift=-0.4cm] (-2,-9.7) node {$\pack(1,8,10)$};

  \edgeAt{0}{-8}{0.3}{-9}{\al}
  \edgeAt{0.3}{-9}{0.6}{-10}{\bl}
 
  \edgeAt{-2.5}{-5}{-2.2}{-6}{\al}

  \fill[white!30!black] (-1.5,-3) circle (0.15cm);

  \fill[white!30!black] (0.3, -9) circle (0.15cm);
  \fill[white!30!black] (0, -8) circle (0.15cm);
  \fill[white!30!black] (0.6, -10) circle (0.15cm);

  \fill[white!30!black] (-2.5,-5) circle (0.15cm);

  \fill[white!30!black] (-2.2,-6) circle (0.15cm);

  \fill[white!30!black] (-1.6,-8) circle (0.15cm);
  
  \fill[white!30!black] (1.2,-5) circle (0.15cm);
\end{tikzpicture}
    \caption{\label{fig:suf_tree_big} Packages from \cref{ex:pack} illustrated as paths (in bold) in the uncompressed suffix trie
      (that is, the trie of all the suffixes) of $w=\al\bl\al\bl\al\al\bl\al\al\bl$.
    }
  \end{center}
  \end{figure}

  \begin{example}\label{ex:pack}
    For a word $w=\al\bl\al\bl\al\al\bl\al\al\bl$,
    a package representation of the set $\Seeds(w)$ is \[\L=\{(1,3,3),(2,9,10),(1,8,10),(3,10,10),(3,7,8),(4,8,8)\}.\]
    It corresponds to the following set $\PACK(\L)$ of all seeds of $w$:
    \begin{center}
      \begin{tabular}{llllll}
        & & $\al\bl\al\bl\al\al\bl\al$ & & & \\
        & $\bl\al\bl\al\al\bl\al\al$ & $\al\bl\al\bl\al\al\bl\al\al$ & & $\al\bl\al\al\bl$ & \\
        $\al\bl\al$ & $\bl\al\bl\al\al\bl\al\al\bl$ & $\al\bl\al\bl\al\al\bl\al\al\bl$ & $\al\bl\al\al\bl\al\al\bl$ & $\al\bl\al\al\bl\al$ & $\bl\al\al\bl\al$ \\
        &&&&& \\
        $\pack(1,3,3)$ & $\pack(2,9,10)$ & $\pack(1,8,10)$ & $\pack(3,10,10)$ & $\pack(3,7,8)$ & $\pack(4,8,8)$
      \end{tabular}
    \end{center}
    \noindent
  This collection of packages is also illustrated in \cref{fig:suf_tree_big} 
  as a set of disjoint paths in the suffix trie of $w$.
\end{example}

  \paragraph{Previous Compact Representation of Seeds.}
  The original linear-size representation of seeds by Iliopoulos, Moore, and Park~\cite{DBLP:journals/algorithmica/IliopoulosMP96},
  also employed in the preliminary version of our work~\cite{DBLP:conf/soda/KociumakaKRRW12},
  requires partitioning the set $\Seeds(w)$ into two disjoint subsets:
  Type-A seeds are represented as paths in the suffix trie of $w$,
  while (the reversals of) type-B seeds admit a similar representation in the suffix trie of the reversed word $w^R$. 
  For both types, the number of reported paths is shown to be linear using a simple argument:
  each path can be uniquely extended to an edge of the corresponding suffix tree (representing subwords of $w$ 
  whose occurrences start at the same positions).

  \begin{remark}\label{rem:repr}
  \cref{ex:pack} illustrated in \cref{fig:suf_tree_big} shows that this argument fails if we try to use it for representing all seeds (rather than just type-A seeds)
  on the suffix tree of $w$: $\al\bl\al\al\bl\al$ and $\al\bl\al\al\bl\al\al\bl$ are both seeds of $w=\al\bl\al\bl\al\al\bl\al\al\bl$
  with the only occurrences starting at position 3, yet they cannot be reported on a single path because $\al\bl\al\al\bl\al\al$ is not a seed of $w$.
  \end{remark}

  \paragraph{Structure of the Paper.}
  We start with a preliminary \cref{sec:prelim}, where we recall several classic notions,
  relate packages to suffix trees, and provide equivalent formal definitions of seeds.
  Next, in \cref{sec:pack}, we prove that seeds admit a package representation of linear size.
  Some of the underlying arguments are stated in an algorithmic way so that they can be used in the subsequent \cref{sec:d-2d}
  as subroutines of a procedure efficiently computing seeds of length $\Theta(n)$.
  In \cref{sec:decomp}, we provide the novel relation between seeds and subword complexity,
  which is the key combinatorial contribution behind our main recursive algorithm described in \cref{sec:main-algorithm}.
  The implementation of an auxiliary operation on package representations
  (applied to merge the results of recursive calls) is deferred to \cref{sec:merge}.

  \paragraph{Our Techniques.}
Our linear-time solution relies on several combinatorial and algorithmic tools.
  \begin{description}
  \item{\sl Compact representation of seeds:}
  Despite its quadratic size and the failure of the naive argument (\cref{rem:repr}),
  the set $\Seeds(w)$ always admits a package representation of linear size (\cref{sec:pack}).
  While this fact is not essential for our algorithm (see \cite{DBLP:conf/soda/KociumakaKRRW12}), it makes our results much simpler and cleaner.
  \item{\sl Combinatorial properties of seeds:}
  The connection between seeds and subword complexity gives an efficient reduction to a set of recursive calls
  of total size decreased by a constant factor (see \cref{sec:decomp,sec:main-algorithm}).
  \item{\sl Interpretation of packages as paths on the suffix trie:}
  Packages naturally correspond to paths in the (uncompacted) suffix trie,
  which can be easily stored using the (compacted) suffix tree. 
  We use this interpretation both to derive the combinatorial upper bound on the package representation size (\cref{sec:pack})
  and in the algorithmic construction of the package representation of long seeds (\cref{sec:d-2d}).
  The new linear-time offline algorithm answering Weighted Ancestor Queries (\cref{subsec:waq}) lets us efficiently map
  packages on the suffix tree. 
  \item{\sl Efficient manipulation of package representations:}
  Package representations provide a convenient way of interpreting the results of recursive calls (seeds of certain subwords)
  as families of subwords of the whole word $w$. This allows for a simple 
  linear-time procedure aggregating the results of the recursive calls (\cref{sec:merge}).
  \end{description}

  \section{Preliminaries}\label{sec:prelim}
  We consider \emph{words} over a polynomially bounded integer alphabet $\Sigma$.
  For a word $w$, by $|w|$ we denote its length and by $w[i]$, for $1\le i \le |w|$,
  we denote its $i$th letter.
  By $\AAlph(w)$ we denote the set of letters occurring in $w$.
For $1\le i \le j \le |w|$, a word $u=w[i] \cdots w[j]$ is called a \emph{subword} of $w$.
We also say that $w$ is a \emph{superstring} of $u$.
In this case, by $w[i\dotdot j]$ we denote the occurrence of $u$ at position~$i$, called a \emph{fragment} of~$w$.
The set of positions where $u$ occurs in $w$ is denoted by $\Occ(u,w)$ (or $\Occ(u)$ if $w$ is clear from the context). 

A fragment of $w$ other than the whole word $w$ is called a \emph{proper} fragment of $w$.
A fragment starting at position $1$ is called a \emph{prefix} of $w$
and a fragment ending at position $|w|$ is called a \emph{suffix} of $w$;
we also say that the corresponding subword is a prefix or suffix of $w$, respectively.

A \emph{border} of $w$ is a subword of $w$ which occurs both as a prefix and as a suffix of $w$.
An integer $p$, $1\le p \le n$, is a \emph{period} of a word $w$ if $w[i]=w[i+p]$ for $1\le i \le |w|-p$.
It is well known that $p$ is a period of $w$ if and only if $w$ has a border of length $|w|-p$; see \cite{AlgorithmsOnStrings,Jewels}.
Moreover, Fine and Wilf's Periodicity Lemma~\cite{fine1965uniqueness} asserts that if a word $w$ has periods $p$ and $q$
such that $p+q-\gcd(p,q) \le |w|$, then $w$ also has a period $\gcd(p,q)$.

Throughtout the paper, by $w$ we denote the word which seeds are to be computed and by $n$ we denote its length.

\subsection{Tries, Suffix Trees, and Package Representations}
\begin{figure}[htpb]
  \begin{center}
    \begin{tikzpicture}[scale=0.7]
  \foreach \x/\y in {0/0,-3/-1,0/-1,0/-2,2/-1, -4/-2,-5/-3,-3/-2,-3/-3,-3/-4,-3/-5,-1.5/-2, -0.5/-3,-1/-4,0.5/-3,1/-4,1.5/-5,2/-6}{
    \filldraw (\x,\y) circle (0.07cm);
  }
  \draw (0,0) -- node[above,sloped] {$\al$} (-3,-1);

  \draw (0,0) -- node[above,sloped] {$\bl$} (0,-1);
  \draw (0,-1) -- node[above,sloped] {$\al$} (0,-2);
  
  \draw (0,0) -- node[above,sloped] {$\mathtt{d}$} (2,-1);
  
  \draw (-3,-1) -- node[above,sloped] {$\al$} (-4,-2);
  \draw (-4,-2) -- node[above,sloped] {$\mathtt{d}$} (-5,-3);

  \draw (-3,-1) -- node[above,sloped] {$\bl$} (-3,-2);
  \draw[line width=.05cm,white!30!black] (-3,-2) -- node[above,sloped] {$\al$} (-3,-3);
  \draw[line width=.05cm,white!30!black] (-3,-3) -- node[above,sloped] {$\al$} (-3,-4);
  \draw[line width=.05cm,white!30!black] (-3,-4) -- node[above,sloped] {$\mathtt{d}$} (-3,-5);

  \draw[decorate, decoration=brace,xshift=-0.3cm] (-3,-5.2) -- (-3,-1.8);

  \draw (-3,-1) -- node[above,sloped] {$\mathtt{d}$} (-1.5,-2);

  \draw (0,-2) -- node[above,sloped] {$\al$} (-0.5,-3);
  \draw (-0.5,-3) -- node[above,sloped] {$\mathtt{d}$} (-1,-4);
  
  \draw (0,-2) -- node[above,sloped] {$\bl$} (0.5,-3);
  \draw[line width=.05cm,white!30!black] (0.5,-3) -- node[below,sloped] {$\al$} (1,-4);
  \draw[line width=.05cm,white!30!black] (1,-4) -- node[below,sloped] {$\al$} (1.5,-5);
  \draw[line width=.05cm,white!30!black] (1.5,-5) -- node[below,sloped] {$\mathtt{d}$} (2,-6);

  \draw[decorate, decoration=brace,xshift=0.4cm,yshift=0.2cm] (0.4,-2.8) -- (2.1,-6.2);

  \draw (-5.5,-4.5) node (A) {equivalence class};
  \draw (-5.5,-5) node {$\pack(2,3,6)$};
  \draw[-latex] (A) -- (-3.5,-3.5);

  \draw (3.5,-2) node {equivalence class};
  \draw (3.5,-2.5) node (B) {$\pack(1,3,6)$};
  \draw[-latex] (B) -- (1.85,-4.2);

  \fill[white!30!black] (-3, -2) circle (.15cm);
  \fill[white!30!black] (-3, -3) circle (.15cm);
  \fill[white!30!black] (-3, -4) circle (.15cm);
  \fill[white!30!black] (-3, -5) circle (.15cm);
  
  \fill[white!30!black] (.5, -3) circle (.15cm);
  \fill[white!30!black] (1, -4) circle (.15cm);
  \fill[white!30!black] (1.5, -5) circle (.15cm);
  \fill[white!30!black] (2, -6) circle (.15cm);

\end{tikzpicture}
    \caption{\label{fig:suf_tree}
      The suffix trie of the word $\texttt{babaad}$.
      The equivalence classes correspond to compacted edges (excluding their topmost nodes).
      Two of them are marked in the figure:
      $\{\al\bl,\al\bl\al,\texttt{abaa},\texttt{abaad}\}=\pack(2,3,6)$ and
      $\{\texttt{bab},\texttt{baba},\texttt{babaa},\texttt{babaad}\}=\pack(1,3,6)$.
    }
  \end{center}
  \end{figure}
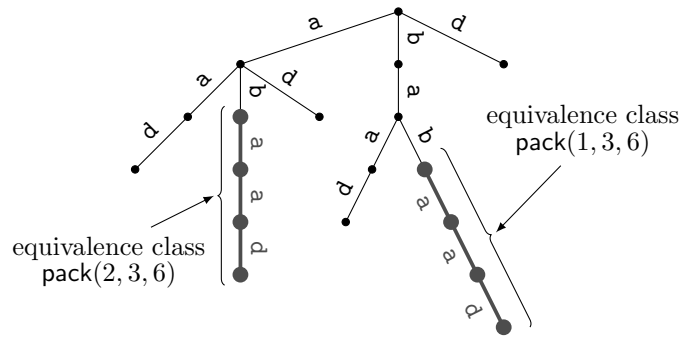

A \emph{trie} is a rooted tree whose nodes correspond to prefixes of words in a given (finite) family $W$.
If $\nu$ is a node, then the corresponding prefix $v$ is called the \emph{value} of the node.
The node with value $v$ is called the \emph{locus} of $v$.
The parent-child relation in the trie is defined so that the root is the locus of the empty word,
while the parent $\mu$ of a node $\nu$ is the locus of the value of $\nu$ with the last character removed.
This character is the \emph{label} of the edge from $\mu$ and $\nu$.
In general, if $\mu$ is an ancestor of $\nu$, then the label of the path from $\mu$ to $\nu$ is the concatenation
of edge labels on the path.
A trie of ithe set $W$ containing all the suffixes of a word \texttt{babaad} is shown in Fig.~\ref{fig:suf_tree}.

A node is \emph{branching} if it has at least two children and \emph{terminal} if its value belongs to~$W$.
A \emph{compacted trie} is obtained from the underlying trie by dissolving all nodes except for the root, branching nodes, and terminal nodes. 
In other words, we compress paths of non-terminal vertices with single children,
and thus the number of remaining nodes becomes bounded by $2|W|$.
We refer to all preserved nodes of the trie as \emph{explicit} (since they are stored explicitly)
and to the dissolved ones as \emph{implicit}.
If $\nu$ is the locus of a word $v$ in an uncompacted trie,
then the locus of $v$ in the corresponding compacted trie is defined as $(\mu, d)$,
where $\mu$ is the lowest explicit descendant of $\nu$, and $d$ is the distance (in the uncompacted trie) from $\nu$ to $\mu$.
Note that $\mu=\nu$ and $d=0$ if $\nu$ is explicit.
Edges of a compacted trie correspond to paths in the underlying trie
and thus their labels are non-empty words, typically stored as references to fragments of the words in $W$.

The \emph{suffix trie} of word $w$ is the trie of all suffixes of $w$ (see Fig.~\ref{fig:suf_tree}), with the locus $w[i \dotdot n]$ labelled by the position $i$.
Consequently, there is a natural bijection between subwords of $w$ and nodes of the suffix trie;
we often use it to identify subwords of $w$ with their loci in the suffix trie.

The \emph{suffix tree} of $w$ \cite{DBLP:conf/focs/Weiner73} is the compacted suffix trie of $w$.
For a word of length $n$, it takes $\Oh(n)$ space and can be constructed in $\Oh(n)$ time
either directly \cite{DBLP:journals/jacm/Farach-ColtonFM00} or from the suffix array of~$w$; see~\cite{DBLP:journals/siamcomp/ManberM93,DBLP:journals/jacm/KarkkainenSB06,Jewels,AlgorithmsOnStrings}.

We say that two subwords $u$ and $v$ of $w$ are \emph{equivalent} if $\Occ(u,w)=\Occ(v,w)$.
The equivalence classes of this relation correspond to edges of the suffix tree of $w$, as shown in the following observation and Fig.~\ref{fig:suf_tree}; see \cite{AlgorithmsOnStrings}.
\begin{observation}
Each equivalence class is of the form
 $E= \pack(i,j_1,j_2)$ for 
some positions $i\le j_1\le j_2$ and corresponds to the set of all nodes on an edge of the suffix tree of $w$ (excluding the topmost explicit note).
Hence, there are at most $2|w|$ equivalence classes.
\end{observation}

  In \cref{sec:merge}, we heavily use the connections between
  suffix trees and package representation to develop the following auxiliary procedure:
  \begin{description}
    \item[$\Combine(R_1,\ldots,R_k)$:]
    Given package representations of sets $R_1,\ldots,R_k$ of subwords of a word $w$,
    compute a smallest package representation of $\bigcap_{i=1}^k R_i$. 
  \end{description}
  Note that the $\Combine$ operation is much more complicated than a simple 
  intersection of sets of intervals.
  The following lemma is proved in \cref{sec:merge}.
  \begin{restatable}{lemma}{lemcombine}\label{lem:combine}
    For a word $w$ of length $n$ and sets $R_1,\ldots,R_k$ of subwords of $w$,
    given in package representations of total size $N$, $\Combine(R_1,\ldots,R_k)$
    can be implemented in $\Oh(n+N)$ time.
    The size of the resulting package representation is at most $N$.
  \end{restatable}

\subsection{Seeds: Formal Definition and Equivalent Characterizations}
  We say that a fragment $w[i\dotdot j]$ \emph{covers} a position $k$ (or that the position $k$ lies within $w[i\dotdot j]$) if $i \le k \le j$.
A word $v$ is  a \emph{cover} of word $w$ if the occurrences of $v$ cover all positions in $w$.
  A word $v$ is a \emph{seed} of word $w$ if it is a subword of $w$ and a cover of a superstring of $w$.
    This definition immediately implies the following observation.
  \begin{observation}\label{obs:hered}
  Let $v$ be a seed of word $w$. If $v$ occurs in a subword $w'$ of $w$, then $v$ is a seed of $w'$.
  \end{observation}
  
  Denote by $\Seeds_I(w)$ the set of all seeds of $w$ with lengths in interval $I$.
  We also denote $\Seeds(w)\,=\, \Seeds_{[1\dotdot n]}(w)$,
  $\Seeds_{\le k}(w)\,=\, \Seeds_{[1 \dotdot k]}(w)$,
  $\Seeds_{\ge k}(w)\,=\, \Seeds_{[k \dotdot n]}(w)$,
  and $\Seeds_k(w)\,=\, \Seeds_{[k\dotdot k]}(w)$.

  A \emph{left-overhanging} occurrence of $v$ in $w$ is a prefix of $w$ that matches a proper suffix of $v$.
  Symmetrically, a \emph{right-overhanging} occurrence of $v$ in $w$ is a suffix of $w$ that matches a proper prefix of $v$.
  The length of the occurrence is the length of the corresponding prefix/suffix of $w$.
  In this context, usual occurrences are sometimes called \emph{full}, while a \emph{generalized} occurrence
  is a full or an overhanging one.
  A more operational definition of seeds can be formulated in terms of generalized occurrences as follows; see \cref{fig:ex_seed}.
  
\begin{fact}[\bf Alternative definition of seeds]\label{fct:overhang}
  A subword $v$ of word $w$ is a seed of $w$ if and only if each position in $w$ is covered by a full, left-overhanging,
  or right-overhanging occurrence of $v$ in $w$.
  \end{fact}
  \begin{proof}
  Suppose that $v$ is a seed of $w$, i.e., a cover of a superstring $xwy$ of $w$. 
  For each position in $w$, the corresponding position in $xwy$ is covered by a full occurrence of $v$ in $xwy$,
  which becomes a generalized occurrence of $v$  when restricted to $w$.
  Hence, generalized occurrences of $v$ in $w$ cover all positions of $w$.
 
 For the converse implication, we construct a superstring $xwy$ of $w$ which has $v$ as a cover.
 For this, we extend $w$ so that the longest left-overhanging occurrence of $v$
 and the longest right-overhanging occurrence of $v$ become full occurrences. 
 Shorter overhanging occurrences may be destroyed, but they do not cover any extra positions in $w$ compared to the longest ones.
 Consequently, the fragment $w$ in $xwy$ is covered by full occurrences of $v$. 
 What is more, the prefix $x$ is covered by the occurrence of $v$ as a prefix of $xwy$ and the suffix $y$ is covered by the occurrence of $v$ as a suffix of $xwy$.
 Thus, $v$ is a cover of $xwy$ and a seed of $w$.
  \end{proof}
  
  We denote by $\F_{k}(w)$ the set all subwords of length $k$ of word 
  $w$. 
\cref{fct:overhang} lets us show that 
the family $\F_{2\ell-1}(w)$ of subwords of length $2\ell-1$ of a word $w$
uniquely determines the length-$\ell$ seeds of $w$.

\begin{lemma}\label{lem:qchar}
  Let $v$ be a non-empty word such that $2|v|-1 \le n$. The following conditions are equivalent:
  \begin{enumerate}[(1)]
    \item\label{it:qchar:a} $v$ is a seed of $w$;
    \item\label{it:qchar:b} $v$ is a seed of every subword of $w$ of length $2|v|-1$.
 \end{enumerate}
\end{lemma}
\begin{proof}
$(\ref{it:qchar:a})\Rightarrow (\ref{it:qchar:b})$.
Each $s\in \F_{2|v|-1}(w)$ occurs as $w[i-|v|+1\dotdot i+|v|-1]$ for some position $i$, $|v|\le i \le n-|v|+1$.
The position $i$ can only be covered by a full occurrence of $v$ contained within this fragment, so $v$ is a subword of each $s\in \F_{2|v|-1}(w)$. Due to \cref{obs:hered}, $v$ is also a seed of every $s \in \F_{2|v|-1}(w)$.

$(\ref{it:qchar:b})\Rightarrow (\ref{it:qchar:a})$.
Positions $i$ satisfying $|v|\le i\le n-|v|+1$ are covered by full occurrences of $v$ due to the fact that
$v$ occurs in each subword of $w$ of length $2|v|-1$, including $w[i-|v|+1\dotdot i+|v|-1]$.
If some position $i < |v|$ or $i > n-|v|+1$ in $w$ is not covered by a full occurrence of $v$, then it can be covered by a left-overhanging occurrence
of $v$ in $w[1\dotdot 2|v|-1]$ or a right-overhanging occurrence of $v$ in  $w[n-2|v|+2\dotdot n]$, respectively.
These overhanging occurrences are also present in $w$.
\end{proof}
 
\begin{corollary}\label{cor:qchar}
For each word $w$ and integer $k$, $1\le k \le \frac{n+1}{2}$, we have
\[\Seeds_{\le k}(w)=\bigcap_{u\in \F_{2k-1}(w)} \Seeds_{\le k}(u).\]
\end{corollary}
\begin{proof}
Consider a length $\ell$ not exceeding $k$.
Note that $\F_{2\ell-1}(w)=\bigcup_{s\in \F_{2k-1}(w)}\F_{2\ell-1}(s),$
so \cref{lem:qchar} yields
\[\Seeds_{\ell}(w)=\bigcap_{u\in \F_{2\ell-1}(w)} \Seeds_{\ell}(u) =\bigcap_{s\in \F_{2k-1}(w)}\bigcap_{u\in \F_{2\ell-1}(s)} \Seeds_{\ell}(u) =\bigcap_{s\in \F_{2k-1}(w)}\, \Seeds_{\ell}(s).\]
The equality holds for each length $\ell\le k$, which concludes the proof.
\end{proof}

  \section{Representation Theorem}\label{sec:pack}

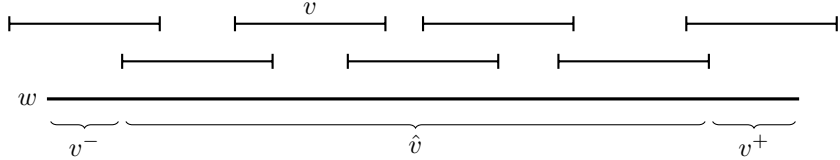
\begin{figure}[ht]
\begin{center}
    \begin{tikzpicture}
    \draw[very thick] (0,0) -- (10,0);
    \draw (0,0) node[left] {$w$};
    \draw[thick,xshift=-0.5cm,yshift=1.0cm] (0,0) -- (2,0)  (0,-0.1) -- (0,0.1)  (2,-0.1) -- (2,0.1);
    \draw[thick,xshift=1cm,yshift=0.5cm] (0,0) -- (2,0)  (0,-0.1) -- (0,0.1)  (2,-0.1) -- (2,0.1);
    \draw[thick,xshift=2.5cm,yshift=1.0cm] (0,0) -- node[above] {$v$} (2,0)  (0,-0.1) -- (0,0.1)  (2,-0.1) -- (2,0.1);
    \draw[thick,xshift=4cm,yshift=0.5cm] (0,0) -- (2,0)  (0,-0.1) -- (0,0.1)  (2,-0.1) -- (2,0.1);
    \draw[thick,xshift=5cm,yshift=1.0cm] (0,0) -- (2,0)  (0,-0.1) -- (0,0.1)  (2,-0.1) -- (2,0.1);
    \draw[thick,xshift=6.8cm,yshift=0.5cm] (0,0) -- (2,0)  (0,-0.1) -- (0,0.1)  (2,-0.1) -- (2,0.1);
    \draw[thick,xshift=8.5cm,yshift=1.0cm] (0,0) -- (2,0)  (0,-0.1) -- (0,0.1)  (2,-0.1) -- (2,0.1);
  
    \draw[decorate, decoration=brace] (9.95,-0.3) -- node[below=0.05cm] {$v^+$} (8.85,-0.3);
    \draw[decorate, decoration=brace] (8.75,-0.3) -- node[below=0.05cm] {$\hat{v}$} (1.05,-0.3);
    \draw[decorate, decoration=brace] (0.95,-0.3) -- node[below=0.05cm] {$v^-$} (0.05,-0.3);
    \end{tikzpicture}
    \end{center}
  \caption{
    Generalized occurrences of a seed $v$ of $w$ and the decomposition of $w$ into $v^-$, $\hat{v}$, and $v^+$. 
    \label{fig:decomp}
  }
  \end{figure}

  Let  $v$ be a subword of a word $w$.
  Let us introduce a decomposition 
  \[w=v^-\cdot  \hat{v}\cdot  v^+\]
   such that $\hat{v}$ 
is the longest subword of $w$ having $v$ as a border.
  In other words, 
  \[\hat{v} = w[\min \Occ(v)\dotdot \max \Occ(v)+|v|-1],\]
  i.e., $\hat{v}$ can be seen as the shortest fragment of $w$ containing all full occurrences of $v$ in $w$;
  see \cref{fig:decomp}.

  \begin{definition}[see \cref{fig:ex_quasi}]
    We say that $v$ is a \emph{quasiseed} of $w$ if it is a cover of $\hat{v}$.
  If $v$ is a seed of $v^-\,v$,
  then $v$ is called a \emph{left candidate} of $w$, and in case 
  $v$ is a seed of $v\,v^+$,
  then $v$ is called a \emph{right candidate}.
  \end{definition}
  By $\Seedseeds(w)$, $\LCands(w)$, and $\RCands(w)$ we denote the sets of quasiseeds, left candidates, and right candidates of $w$, respectively.
  
  \begin{lemma}\label{lem:seed}
    $\Seeds(w) = \LCands(w) \cap \Seedseeds(w) \cap \RCands(w)$.
  \end{lemma}
  \begin{proof}
  We apply the characterization of \cref{fct:overhang}.
  Observe that there are natural bijections between full occurrences of $v$ in $w$ and in $\hat{v}$,
  between left-overhanging occurrences of $v$ in $w$ and in $v^-v$,
  and between right-overhanging occurrences of $v$ in $w$ and in $vv^+$.
  
  Next, note that any position of $w$ within $\hat{v}$ covered an overhanging occurrence of $v$
  must be located within the leading or trailing $|v|$ characters of $\hat{v}$, so it is also covered by a full occurrence of $v$ since $v$ is a border of $\hat{v}$.
  Thus, $v$ is a quasiseed of $w$ if and only if the positions of $w$ within $\hat{v}$ are covered by generalized occurrences of~$v$.
  
  Moreover, observe that $v$ occurs in $v^-v$ only as a suffix,  so the positions within the leading $v^-$ of $v^-v$ and of $w$ can be covered by left-overhanging occurrences only. Consequently, $v$ is a left candidate if and only if the positions within the leading $v^-$ of $w$ are covered by generalized occurrences of $v$.
  Symmetrically, $v$ is a right candidate if and only if the positions within the trailing $v^+$ are covered by generalized occurrences of~$v$.
  
Combining these three facts, we conclude that $v$ is a seed of $w$ if and only if it is simultaneously a quasiseed, a left candidate, and a right candidate.
  \end{proof}
  \begin{figure}[t]
      \begin{center}
    \begin{tikzpicture}[scale=0.7]
\clip (-0.2,-0.41) rectangle (8.7,0.96);
\fill[black!15] (-0.2,0) rectangle (0.7,0.6)  (1.8,0) rectangle (7.2,0.6)  (8.3,0) rectangle (8.8,0.6);
\foreach \x in {0,1,3,4,6,8,9,11,13,14,16,17} {
\draw (\x/2,0) node [above] {{$\al$}};
}
\foreach \x in {2,5,7,10,12,15} {
\draw (\x/2,0) node [above] {{$\bl$}};
}
\
\foreach \x in {-4,9} {
   \draw[thick,yshift=0.15cm] (\x/2-0.2,0.5) -- (\x/2-0.2,0.6) -- (\x/2+2.7,0.6) -- (\x/2+2.7,0.5);
}
\foreach \x in {4,17} {
   \draw[thick] (\x/2-0.2,0) -- (\x/2-0.2,-0.1) -- (\x/2+2.7,-0.1) -- (\x/2+2.7,0);
}
\end{tikzpicture}
  \end{center}
  \begin{center}
    \begin{tikzpicture}[scale=0.7]
\clip (-0.2,-0.41) rectangle (8.7,0.96);
\fill[black!15] (-0.2,0) rectangle (2.7,0.6)  (3.3,0) rectangle (5.2,0.6)  (5.8,0) rectangle (8.8,0.6);
\foreach \x in {0,1,3,4,6,8,9,11,13,14,16,17} {
\draw (\x/2,0) node [above] {{$\al$}};
}
\foreach \x in {2,5,7,10,12,15} {
\draw (\x/2,0) node [above] {{$\bl$}};
}
\
\foreach \x in {-1,7,15} {
   \draw[thick,yshift=0.15cm] (\x/2-0.2,0.5) -- (\x/2-0.2,0.6) -- (\x/2+1.7,0.6) -- (\x/2+1.7,0.5);
}
\foreach \x in {2,12} {
   \draw[thick] (\x/2-0.2,0) -- (\x/2-0.2,-0.1) -- (\x/2+1.7,-0.1) -- (\x/2+1.7,0);
}
\end{tikzpicture}
  \end{center}
    \caption{\label{fig:ex_quasi}
        The word $\texttt{ababaa}$ is a quasiseed of $w=\texttt{aabaababaababaabaa}$,
        whereas the word $\texttt{baab}$ is both a left and a right candidate of $w$.
        Neither of them is a seed of $w$, though.
    }
  \end{figure}
    Next, we shall characterize the quasiseeds and candidates in a computationally feasible way
    and bound the sizes of their package representations.

  \subsection{Quasiseeds}

 For a set $X=\{x_1,\ldots,x_k\}$ of integers, $x_1 < \cdots < x_k$, let us define the value $\maxgap(X)$ as:
  \[
    \maxgap(X)\ =\begin{cases}
    0 & \text{ if }k \le 1,\\
    \max\{x_{i+1}-x_{i} : 1 \le i < k\} & \text{ if }k \ge 2.
    \end{cases}
  \]
  For example, $\maxgap(\{1,3,8,13,17\})=5$.
  The following easy observation relates this function to covers; see, e.g., \cite{DBLP:journals/iandc/BerkmanIP95}.

%
%
%

\begin{observation}\label{obs:char:qs}
A subword $v$ of word $w$ is a quasiseed of $w$ if and only if $\maxgap(\Occ(v,w))\le |v|$.
\end{observation}
%
%

\begin{corollary}
For each equivalence class $E$, the quasiseeds contained in $E$ form a single package. In other words,
$\Seedseeds(w)\cap E$ has a package representation of size 1.
\end{corollary}
\begin{proof}
Consider the shortest subword $v\in E\cap \Seedseeds(w)$ and let $u\in E$ satisfy $|u|\ge |v|$.
Due to \cref{obs:char:qs}, $\maxgap(\Occ(u))=\maxgap(\Occ(v))\le |v|\le |u|$, 
so $u$ is also a quasiseed. This completes the proof.
\end{proof}

\subsection{Left Candidates}\label{subsec:lc}
The \emph{border table} $B[0\dotdot n]$ stores at $B[i]$ the length of the longest proper border of $w[1\dotdot i]$
(we assume $B[0]=-1$).
The following fact was already shown implicitly in~\cite{DBLP:journals/tcs/ChristouCIKPRRSW13}; we give its proof for completeness.

\begin{lemma}\label{lem:char:lc}
A subword $v$ of word $w$ is a left candidate of $w$ if and only if $B[|v^-v|]\ge |v^-|$.
\end{lemma}
\begin{proof}
Recall that positions within the prefix $v^-$ of $v^-v$ can only be covered
by left-overhanging occurrences of $v$.

Consequently, if $v$ is a seed of $v^-v$, then there is a prefix of $v^-v$ of length at least $|v^-|$
equal to a suffix of $v$. In other words, there is a proper border of $v^-v$ of length at least $|v^-|$.

For a proof of the converse implication, suppose that $v^-v$ has a proper border
of length at least $|v^-|$. Since $v$ has only one occurrence in $v^-v$, the border must be a proper suffix of $v$.
Consequently, $v$ has a left-overhanging occurrence in $v^-v$ of length at least $|v^-|$. This fragment
covers all positions in $v^-$; the remaining positions in $v^-v$ are trivially covered by the occurrence of $v$ as a suffix of $v^-v$.
\end{proof}

A classic property of the border table is that $0\le B[i]\le B[i-1]+1$ holds for $1 \le i \le n$.
Fine and Wilf's Periodicity Lemma~\cite{fine1965uniqueness} lets us further characterize positions where the right inequality is strict.
  \begin{lemma}\label{lem:per}
If $B[i]\le B[i-1]$ holds for a position $i$ of the word $w$, then $B[i]+B[i-1] < i-1$.
  \end{lemma}
\begin{proof}
We assume $B[i]>0$; otherwise, the claim holds trivially: $B[i]+B[i-1]=B[i-1]< i-1$.
Since $w[1\dotdot i]$ does not have a border of length $B[i-1]+1$, we must have
$w[B[i-1]+1]\ne w[i]=w[B[i]]$, so $B[i-1]+1-B[i]$ is not a period of $w[1\dotdot i-1]$
despite the fact that both $i-B[i]$ and $i-1-B[i-1]$ are periods of this prefix.
By the Periodicity Lemma, this yields $(i-B[i])+(i-1-B[i-1])-1>i-1$,
so $B[i]+B[i-1]<i-1$ holds as claimed.
\end{proof}

\newcommand{\ACTIVE}{\mathsf{Active}}

 \newcommand{\NSV}{\mathcal{N}}
 Let $v$ be a subword of $w$ and let $v=v'a$ where $a\in \Sigma$.
 We say that $v$ is a \emph{critical left candidate} if it is a left candidate, but $v'$ is not a left candidate (in particular, $v'$ can be the empty word)
 or $\min \Occ(v') < \min \Occ(v)$.
 Let
 \[\ACTIVE(w)\;=\; \{\, j\in [1\dotdot n]\ :\ B[j-1] < B[j]\le \tfrac{j-1}{2}\,\}.\] 
 The following lemma is the main technical contribution in the characterization of left candidates.

 \pagebreak
 \begin{lemma}\label{lem:lcand}
A fragment $w[i\dotdot j]$ is
   \begin{enumerate}[(a)]
     \item\label{it:lcand:a} the leftmost occurrence of a left candidate if and only if $i-1 \le B[j] \le j-i$;
      \item\label{it:lcand:b} the leftmost occurrence of a critical left candidate if and only if
      $i=B[j]+1$ and $j\in \ACTIVE(w)$.
   \end{enumerate}
 \end{lemma}
 \begin{proof}
\mbox{ \ }\\
   \eqref{it:lcand:a} 
   If $w[i\dotdot j]$ is the leftmost occurrence of a left candidate $v$, then $w[1\dotdot j]=v^-v$ and $B[j]\ge |v^-|=i-1$ by \cref{lem:char:lc}.
   Moreover, $B[j]<|v|=j-i+1$, because $w[i-j+B[j]\dotdot B[j]]$ would otherwise be an earlier occurrence of~$v$.
   These two conditions yield $i-1 \le B[j] \le j-i$.

    Conversely, assume that
    \begin{equation}\label{eq:00}
      i-1\le B[j] \le j-i.
    \end{equation}

   Suppose for a proof by contradiction that $v=w[i\dotdot j]$ has an earlier occurrence at position $i'$,
   i.e., $w[i'\dotdot i'+j-i]=w[i\dotdot j]$ for some position $i'<i$.
   Let $w'=w[i' \dotdot j]$.
   The word $v$ is a border of $w'$, so $w'$ has a period $i-i'$, which we denote by $p_1$.
   The shortest period of the whole prefix $w[1 \dotdot j]$, hence a period of $w'$, is $j-B[j]$, which we denote by $p_2$.
   By the first inequality of \eqref{eq:00},
   $$p_1+p_2 \,=\, i-i' + j-B[j] \,\le\, j-i'+1 \,=\, |w'|.$$
   Hence, $p_1$ and $p_2$ satisfy the assumption of the Periodicity Lemma and $w'$ has period $\gcd(p_1,p_2)$, which we denote by $p$.
   Moreover, by the second inequality of \eqref{eq:00},
   $$p_1 \,=\, i-i' < i \,\le\, j-B[j] \,=\, p_2,$$
   so $p<p_2$ and $p$ divides $p_2$.
   Consequently, $w[1 \dotdot j]$ has period $p$, which contradicts the fact that $p_2$ is its shortest period.

   Therefore, $w[i\dotdot j]$ indeed is the leftmost occurrence of $v$ and $v^-v = w[1\dotdot j]$.
   Due to $B[j]\ge i-1=|v^-|$, we conclude that $v$ is a left candidate by \cref{lem:char:lc}.
  
\medskip
   \noindent
   \eqref{it:lcand:b}
   First, assume that 
{$B[j-1]<B[j]\le\frac{j-1}{2}$ and $i=B[j]+1$.}

   Note that 
   \[i-1=B[j]=2B[j]-B[j]\le j-1-B[j]=j-i.\]
   By part~\eqref{it:lcand:a}, $w[i\dotdot j]$ is therefore the leftmost occurrence of a left candidate.
   On the other hand, $B[j-1]<B[j]=i-1$, so $w[i\dotdot j-1]$ is not the leftmost occurrence of a left candidate.
   Thus, $w[i\dotdot j]$ is the leftmost occurrence of a critical left candidate.
  
   Next, assume that $w[i\dotdot j]$ is the leftmost occurrence of a critical left candidate. 
   Part~\eqref{it:lcand:a} yields $i-1\le B[j]\le j-i$ (since $w[i\dotdot j]$ is the leftmost occurrence of a left candidate)
   and that $B[j-1]< i-1$ or $B[j-1]\ge j-i$ (since $w[i\dotdot j-1]$ is not).
  
   In the latter case, we have $B[j-1]+B[j]\ge j-i+i-1= j-1$, so $B[j]=B[j-1]+1$ holds by \cref{lem:per}.
   This yields a contradiction:
   \[j-i\ge B[j]=B[j-1]+1 \ge j-i+1.\]

   \noindent Thus, the only possibility is that $B[j-1]<i-1$, i.e., $B[j-1]\le i-2$.
   We then have 
   \[B[j]\le B[j-1]+1 \le i-1 \le B[j],\] 
   so $B[j-1]<B[j]=i-1$.
   Furthermore, 
   \[B[j]= \tfrac12(B[j]+B[j]) \le \tfrac12(j-i+i-1)=\tfrac{j-1}{2}\]  
   holds as claimed.
 \end{proof}

\noindent  For a table $A[0\dotdot n+1]$, assuming $A[n+1] = -\infty$,
define the \emph{nearest smallest value} table $\NSV_A$ such  that
for $0\le i\le n$ we have:  $\NSV_A[i] =\min \{j> i : A[j]< A[i]\}.$

\begin{lemma}\label{cor:lcand}
Packages $\pack(B[j]+1,j,\NSV_B[j]-1)$ {for} $j\in \ACTIVE(w)$
form a package representation of the family of left candidates of $w$.
This representation (of size at most $n$) can be computed in $\Oh(n)$ time.
\end{lemma}
 \begin{proof}
First, we shall prove that for each $j\in \ACTIVE(w)$, the fragments $w[B[j]+1\dotdot k]$ for $j \le k \le \NSV_B[j]-1$
are leftmost occurrences of left candidates.
To apply the characterization of \cref{lem:lcand}\eqref{it:lcand:a},
we need to show that $B[j]\le B[k]\le k-B[j]-1$.
The first inequality follows directly from the definition of the $\NSV_B$ table,
while for the second one, we observe that $B[j]+B[k]\le 2B[j]+k-j\le j-1+k-j = k-1$ holds due to $B[j]\le\frac{j-1}{2}$
and the classic property of the border table.

Consequently, the packages are disjoint and consist of left candidates only.
 
It remains to prove that we do not leave any left candidate behind.
Suppose that $w[i\dotdot k]$ is the leftmost occurrence of a left candidate.
Repeatedly trimming the trailing character, we can reach a critical left candidate whose leftmost occurrence is $w[i\dotdot j]$
(for $j\le k$).
By \cref{lem:lcand}\eqref{it:lcand:b}, we have that $j\in \ACTIVE(w)$ and $i=B[j]+1$.
However, since $w[i\dotdot k']$ is the leftmost occurrence of a left candidate for all $j \le k' \le k$,
\cref{lem:lcand}\eqref{it:lcand:a} yields $B[k']\ge i-1 = B[j]$ for $j \le k' \le k$.
Consequently, $\NSV_B[j]> k$.
Thus, $w[i\dotdot k]\in \pack(B[j]+1,j, \NSV_B[j]-1)$ indeed belongs to one of the packages we created. 
 
The size of the package representation is $|\ACTIVE(w)|\le n$.
As for the $\Oh(n)$-time construction algorithm, we first build the border table $B$~\cite{morris1970linear,AlgorithmsOnStrings,Jewels} and its nearest smallest value table $\NSV_B$ 
(using a Cartesian tree construction algorithm~\cite{DBLP:conf/stoc/GabowBT84}).
Then, for each position $j$, we test in constant time whether $j\in \ACTIVE(w)$ and, if so, we retrieve the corresponding package.
\end{proof}

\subsection{Right Candidates}
 For word $w$, let us define a \emph{reverse border array} $B^R[1\dotdot n]$ such that $B^R[i]$ is the length of the longest proper border
 of $w[i\dotdot n]$.
 
\begin{lemma}\label{lem:char:rc}
A subword $v$ of a word $w$ is a right candidate of $w$ if and only if $B^R[n-|vv^+|+1]\ge |v^+|$.
 \end{lemma}
 \begin{proof}
Follows from \cref{lem:char:lc} by the symmetry between $B$ and $B^R$ as well as left and right candidates.
\end{proof}

Even though \cref{lem:char:rc} for right candidates and \cref{lem:char:lc} for left candidates are symmetric, the former allows us to represent right candidates
on each edge of the suffix tree of $w$ as a single package.
Thus a representation for right candidates is much simpler to be computed than the representation for left candidates.

\begin{lemma}\label{lem:rcand}
The intersection of $\RCands(w)$ with a single equivalence class 
forms at most one package.
Moreover, a package representation (of size at most $2n$) of 
the set $\RCands(w)$ can be computed in $\Oh(n)$ time.
\end{lemma}
\begin{proof}
Let us consider an equivalence class $E$ of subwords of $w$, with $P=\Occ(v)$ for $v \in E$, and denote
\[\ell(E)= n-\max P+1-B^R[\max P].\]
We will show that $v\in E$ is a right candidate if and only if $|v| \ge \ell(E)$.
By \cref{lem:char:rc}, $v$ is a right candidate if and only if $B^R[n-|vv^+|+1]\ge |v^+|$.
However, $|vv^+|=n-\max P + 1$ and  $|v^+| = n-\max P +1 - |v|$, so the two conditions are equivalent.

Let us show how to compute $\RCands(w)\cap E$
for each equivalence class $E=\pack(i,j_1,j_2)$.
We construct the array $B^R$~\cite{morris1970linear,AlgorithmsOnStrings,Jewels} and the suffix tree of $w$.
Moreover, for each equivalence class $E$, we determine the common value $\max(\Occ(v))$ for $v\in E$.
This lets us compute $\ell(E)$.
We have proved that the right candidates in $E$ are precisely
words in $E$ of length $\ell(E)$ or more. If $\ell(E)>|w[i\dotdot j_2]|$, there are no right candidates in $E$.
Otherwise, we output a package \[\pack(i, \max(j_1, i+\ell(E)-1), j_2).\qedhere\]
\end{proof}

\subsection{Representation Theorem for Seeds}

  \begin{theorem}\label{thm:comb}
  For a word $w$ of length $n$, the set $\Seeds(w)$ has a package representation of size at most $3n$.
  \end{theorem}
\begin{proof}
By \cref{lem:seed}, we have $\Seeds(w)=\LCands(w)\cap \Seedseeds(w)\cap \RCands(w)$.

First, we shall prove the package representation of $\Seedseeds(w)\cap \RCands(w)$
consists of at most $2n$ packages. 
Indeed, for each equivalence class $E$, both $\Seedseeds(w)\cap E$
and $\RCands(w)\cap E$ form at most one package. The intersection of two packages
forms at most one package, so $\Seedseeds(w)\cap \RCands(w)$
has a package representation with at most one package per equivalence class,
i.e., of total size at most $2n$.

By \cref{cor:lcand}, $\LCands(w)$ has a package representation of size at most $n$.

Finally, \cref{lem:combine} implies that $\Seeds(w)$ has a package representation of size at most $3n$.
\end{proof}

   \section{Computing Long Seeds}\label{sec:d-2d}
   In this section, we provide a linear-time algorithm computing seeds of length $\Theta(n)$. 
   Formally, we consider the following operation for $\ell=\Omega(n)$.
   \begin{description}
    \item[$\LSeedsRep(\ell,w)$:]
    Computes a package representation of $\Seeds_{\ge \ell}(w)$.
\end{description}
   Let us denote by $\Seedseeds_{\ge \ell}(w)$ the set of all quasiseeds of $w$ with lengths at least $\ell$.
   Our implementation is based on \cref{thm:comb}, and it uses the suffix tree of $w$ to determine $\Seedseeds_{\ge \ell}(w)$. 

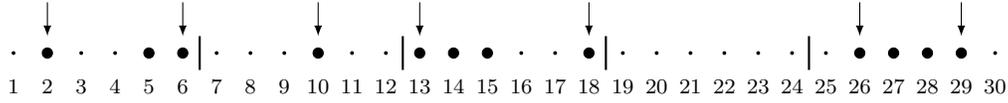
\begin{figure}[htpb]
    \begin{center}
\begin{tikzpicture}[scale=0.45]
\foreach \x in {1,...,30} \filldraw (\x,0) circle (0.04cm);
\foreach \x in {1,...,30} \draw (\x,-1) node {\footnotesize \x};
\foreach \x in {2,5,6,10,13,14,15,18,26,27,28,29} \filldraw (\x,0) circle (0.15cm);
\foreach \x in {2,6, 10, 13,18, 26,29} \draw[xshift=\x cm,-latex] (0,1.5) -- (0,0.5);
\foreach \x in {6,12,18,24} \draw[thick,xshift=0.5cm] (\x,-0.5) -- (\x,0.5);
\end{tikzpicture}
\end{center}
    \vspace*{-0.4cm}
    \caption{\label{fig:ex_refine}
      For $n=30$ and $\ell=6$,
      $\refine(\{2,5,6,10,13,14,15,18,26,27,28,29\})\,=\,\{2,6, 10, 13,18, 26,29\}$.
    }
  \end{figure}

   Let us partition the positions of $w$ into a family $\Fa$ of $\Oh(n/\ell)$ disjoint \emph{blocks}
   of length at most $\ell$ each. 
   For a set of positions $X$, we define its refined version:
   \[\refine(X)\;=\; \bigcup_{Y\in \Fa \, : \, Y\cap X \ne \emptyset}  \{\min(Y\cap X),\, \max(Y\cap X)\};\]
   see Fig.~\ref{fig:ex_refine}.
   Note that $|\refine(X)| \le 2|\Fa| = \Oh(n/\ell) = \Oh(1)$.

      \begin{lemma}\label{lem:refine}
   A subword $v$ of length at least $\ell$
   is a quasiseed if and only if $\maxgap(\refine(\Occ(v)))\le |v|$.
   \end{lemma}
   \begin{proof}
   Clearly, $\refine(\Occ(v))\sub\Occ(v)$, so $\maxgap(\refine(\Occ(v)))\ge \maxgap(\Occ(v))$. Due to \cref{obs:char:qs},
   it remains to prove that $\maxgap(\refine(\Occ(v)))> |v|$ yields $\maxgap( \Occ(v))> |v|$.
   Let $i<i'$ be consecutive elements of $\refine(\Occ(v))$ such that $i'-i>|v|$.
   Since $|v|\ge \ell$, positions $i$ and $i'$ belong to different blocks of $\Fa$.
   Moreover, these positions are consecutive elements of $\refine(\Occ(v))$,
   so $i$ must be the largest in its block, $i'$ must be the smallest in its block, and all blocks in between cannot contain any element of $\Occ(v)$.
   Consequently, $i<i'$ are consecutive elements of $\Occ(v)$, i.e., $\maxgap( \Occ(v))> |v|$.
   \end{proof}

   We traverse the suffix tree of $w$ bottom-up, computing $\refine(\Occ(v))$ in constant time
   for each subword $v$ whose locus is an explicit node. 
   \begin{fact}\label{fct:refine}
    The sets $\refine(\Occ(v))$ for all explicit nodes $v$
   of the suffix tree can be computed in $\Oh(n)$ time.
     \end{fact}
     \begin{proof}
     We compute $\refine(\Occ(v))$ for each explicit node $v$ and store it as a sorted list.
     For this, we start with an empty set and consider all explicit children $u$ of $v$.
     For each child, we merge the current list with $\refine(\Occ(u))$, removing elements which are not extremes in their blocks.
     This takes $\Oh(|\Fa|)=\Oh(1)$ time for each edge of the suffix tree, which gives $\Oh(n)$ time in total.
     \end{proof}

     \begin{lemma}[\bf $\LSeedsRep$ Implementation]\label{lem:long}
      For a threshold $\ell=\Theta(n)$ and a word $w$ of length $n$, $\LSeedsRep(\ell,w)$
      can be implemented in $\Oh(n)$ time.
    \end{lemma}
     \begin{proof}  
     First, we compute a package representation of the family $\Seedseeds_{\ge \ell}(w)$ of long quasiseeds.
     Consider an equivalence class $E=\pack(i, j_1, j_2)$.
     Let $P$ be the common occurrence set $\Occ(v)$ for $v\in E$.
     The longest subword in each package is represented by an explicit node,
     so the procedure of \cref{fct:refine} computes $\refine(P)$.
     Let us define $\ell' := \max(\ell, \maxgap(\refine(P)))$.
     By \cref{lem:refine}, a subword $v\in E$ is a long quasiseed if and only if $|v|\ge \ell'$.
     If $j_2 - i +1 < \ell'$, there are no such quasiseeds.
     Otherwise, we report a package
    \[E \cap \Seedseeds_{\ge \ell}(w) = \pack(i, \max(j_1,i+\ell'-1), j_2).\]
     Finally, we apply \cref{lem:seed} and compute $\Seeds_{\ge \ell}(w)=\Seedseeds_{\ge \ell}(w)\cap \LCands(w)\cap\RCands(w)$ using \cref{lem:combine}
     to implement the intersection. Linear-size package representations of left candidates and right candidates
     are constructed using \cref{cor:lcand,lem:rcand}, respectively.
     The total running time is $\Oh(n)$.
\end{proof}

We also use the following operation that can be computed using \LSeedsRep.

\begin{description}
    \item[$\BSeedsRep(I,w)$:]
    Computes a package representation of $\Seeds_I(w)$.
\end{description}

Our implementation of $\BSeedsRep(I,w)$ uses a reduction to \cref{lem:long,lem:combine},
and its running time is linear provided that $I$ is \emph{balanced}, i.e., if the ratio of its endpoints in bounded by a constant.
 \begin{lemma}[\bf $\BSeedsRep$ Implementation]\label{lem:qseedsrep}
    For an interval $I=[\ell\dotdot r]$ and a word $w$ of length $n$, $\BSeedsRep(I,w)$
    can be implemented in $\Oh(n)$ time if $r=\Oh(\ell)$.
  \end{lemma}

\begin{proof} 
   We construct a family $R$ of fragments of length $4r$ covering $w$ with overlaps of size $2(r-1)$
   (the last fragment might be shorter). Note that the total length of these fragments is at most $2n$.
   Furthermore, observe that $\F_{2r-1}(w)=\bigcup_{s\in R} \F_{2r-1}(s)$, so \cref{cor:qchar}
   yields 
   \[\Seeds_{\le r}(w)=\bigcap_{u\in \F_{2r-1}(w)} \Seeds_{\le r}(u) = \bigcap_{s\in R}\;\bigcap_{u\in \F_{2r-1}(s)}\Seeds_{\le r}(u)=\bigcap_{s\in R}\Seeds_{\le r}(s).\]
    In particular, $\Seeds_{I}(w)=\bigcap_{s\in R}\Seeds_{I}(s)$.
   For each $s\in R$, we apply \cref{lem:long} to determine a package representation of $\Seeds_{\ge \ell}(s)$, and we filter out seeds of length greater than $r$.
   This takes $\Oh(|s|)$ time for each $s\in R$, which is $\Oh(n)$ in total.
   Finally, we combine the package representations in $\Oh(n)$ time using \cref{lem:combine}.
   \end{proof}

\section{Relation between Seeds, Compression and Subword Complexity}\label{sec:decomp}
The \emph{subword complexity} of a word $w$ is a function which gives the number of subwords 
of a given length $k$, i.e., $|\F_{k}(w)|$. 
Since $|\F_k(w)|$ is not monotone in general, 
we define a non-decreasing sequence $(\beta_k)_{k=1}^n$ with \[\beta_k(w) = |\F_{k}(w)|+k-1;\]
see Fig.~\ref{tab:F_beta}.

\begin{figure}[htpb]
  \centering
  \begin{tabular}{r|c|c|c|c|c|c|c|c}
    $k$ & 1 & 2 & 3 & 4 & 5 & 6 & 7 & 8 \\\hline
    $w[k]$ & \texttt{a} & \texttt{b} & \texttt{a} & \texttt{b} & \texttt{a} & \texttt{a} & \texttt{b} & \texttt{b} \\\hline
    $|\F_k(w)|$ & 2 & 4 & 5 & 5 & 4 & 3 & 2 & 1 \\\hline
    $\beta_k(w)$ & 2 & 5 & 7 & 8 & 8 & 8 & 8 & 8
  \end{tabular}
  \caption{Subword complexity of $w$ and $\beta_k(w)$ for $w=\mathtt{ababaabb}$.}
  \label{tab:F_beta}
\end{figure}

A connection between the subword complexity and the existence of seeds of certain lengths is crucial in the paper.
More precisely, each seed provides an upper bound on the values $\beta_k(w)$.
\begin{lemma}[\bf Gap Lemma]\label{lem:mks}
If $\beta_k(w)> \frac{2}{3}n$, then $w$ has no seed $v$ whose length
satisfies $2k-2 \le |v| \le \frac16n$.
\end{lemma}
\begin{proof}
We shall prove that if $v$ is a seed of $w$, then $\beta_k(w) \le |v| + (k-1)\frac{n}{|v|}$.

Note that we may assume $|v|\ge k$; 
otherwise, the right-hand side is at least $n$.
Consider length-$k$ fragments starting at positions $i,\ldots,i+\ell$, where $\ell < |v|$.
We claim that at most $k-1$ of these fragments are not covered by single occurrences of $v$.
Let $w[i'\dotdot i'+k-1]$, for $i \le i' \le i+\ell$, be the first such fragment that is not covered by any single occurrence of $v$.
If $i'$ does not exist or $i'+k-1 > i+\ell$, we are done.
Otherwise, the occurrence of $v$ covering position $i'+k-1$ must start at some position $j$, $i' < j < i'+k$.
Consequently, the length-$k$ fragments starting at positions $i''$, $j \le i'' \le j+|v|-k$,
are all covered by this occurrence of $v$.
We are left with at most $k-1$ remaining length-$k$ fragments (starting at positions $i''$ such that $i'\le i'' < j$ or $j+|v|-k < i'' \le i+\ell$).

Thus, at most $(k-1)\big\lceil{\frac{n-|v|+1}{|v|}}\big\rceil$ length-$k$ fragments of $w$ are not covered by single occurrences of $v$.
As a result, $|\F_k(w)\setminus \F_k(v)|\le (k-1)\big\lceil{\frac{n-|v|+1}{|v|}}\big\rceil$, and we obtain the claimed the upper bound on $\beta_k(w)$:
\[
\beta_k(w) = k-1+|\F_k(w)| = k-1+|\F_k(v)|+|\F_k(w)\setminus \F_k(v)| \le 
|v|+ (k-1)\left\lceil{\tfrac{n-|v|+1}{|v|}}\right\rceil \le |v|+(k-1){\tfrac{n}{|v|}}.
\]
If $2k-2 \le |v| \le \frac16 n$, then we conclude that 
\[\beta_k(w) \le |v| + (k-1)\tfrac{n}{|v|}\le \tfrac16 n + \tfrac12 n=\tfrac{2}{3}n,\]
which contradicts our assumption.
\end{proof}

\cref{lem:mks} yields a gap in the feasible lengths of seeds of $w$.
Seeds of length $\frac16n$ or more can be determined using the \LSeedsRep procedure,
so we may focus on computing relatively short seeds. 
Due to the characterization of \cref{lem:qchar}, we may ignore some regions of $w$ as long as we do not miss
any subword of certain length.
To formalize this intuition, we define the following operation $\Compress_k(w)$,
which results in a set of subwords $S$ of $w$
such that $\F_k(w)=\bigcup_{s\in S} \F_k(s)$.

Let us take the set of fragments which are leftmost occurrences of subwords in $\F_k(w)$. 
Any two overlapping or consecutive fragments are joined together, and the subwords indicated by the resulting set
of fragments give the family $\Compress_k(w)$; see \cref{blocks} for an example.

\begin{figure}[t]
\centering
  \begin{tikzpicture}[scale=0.4]
    \begin{scope}[yshift=0cm]
    \foreach \l/\r in {0/3, 6/9, 15/20, 21/24,28/30} {
      \filldraw[black!20] (\l+0.5,0) rectangle (\r+0.5,1);
    }
    \draw[thick] (0.5,0) rectangle (30.5,1);
    \end{scope}
    \foreach[count=\x] \c in {a,b,a,b,a,b,a,c,a,c,a,c,a,c,a,c,c,b,a,a,c,b,b,c,b,c,a,b,a,d}{
      \draw (\x,0) node[above=-.03] {$\mathtt{\c}$};
    }
    \foreach \x in {1,5,10,15,20,25,30}{
      \draw (\x,1) node[above]{\tiny \x};
    }
  \end{tikzpicture}
\vspace*{-0.4cm}
\caption{
  $\Compress_2(w)=
  \{w[1\dotdot 3], w[7\dotdot 9], w[16\dotdot 20], w[22\dotdot 24], w[29\dotdot30]\}
  = \{\mathtt{aba},\mathtt{aca},\mathtt{ccbaa},\mathtt{bbc},\mathtt{ad}\}$.
  We have $\F_{2}(\mathtt{aba}) \cup \F_{2}(\mathtt{aca})\cup \F_{2}(\mathtt{ccbaa})\cup \F_{2}(\mathtt{bbc})\cup \F_2(\mathtt{ad})
  =\{\mathtt{ab},\mathtt{ba}\}\cup \{\mathtt{ac,ca}\}\cup \{\mathtt{cc,cb,ba,aa}\} \cup \{\mathtt{bb,bc}\}\cup \{\mathtt{ad}\}
  =\F_{2}(w) .$
}\label{blocks}
\end{figure}

Recall that our motivation is to reduce computing short seeds in $w$ to the analogous operation in each $s\in \Compress_k(w)$.
This is illustrated by the following result.
\begin{lemma}[\bf Reduction Lemma]\label{lem:red}\mbox{ \ }\\
Consider a word $w$ of length $n$. 
For every integer $k$, $1\le k\le \frac{n+1}{2}$, we have 
\[\Seeds_{\le k}(w) = \bigcap_{s\in \Compress_{2k-1}(w)}\, \Seeds_{\le k}(s).\]
\end{lemma}
\begin{proof}
Note that $\F_{2k-1}(w)=\bigcup_{s\in \Compress_{2k-1}(w)}\F_{2k-1}(s).$
Hence, \cref{cor:qchar} yields
\[\Seeds_{\le k}(w)=\!\!\!\bigcap_{u\in \F_{2k-1}(w)} \Seeds_{\le k}(u) =\!\!\!\bigcap_{s\in \Compress_{2k-1}(w)}\bigcap_{u\in \F_{2k-1}(s)} \Seeds_{\le k}(u) =\!\!\!\bigcap_{s\in \Compress_{2k-1}(w)}\, \Seeds_{\le k}(s).\]
This concludes the proof.
\end{proof}
The values $\beta_k(w)$ can be used to bound the total length of words $s\in \Compress_k(w)$,
denoted $\|\Compress_k(w)\|$.

\begin{lemma}[\bf Compression Lemma]\label{lem:mkbk}
For each word $w$ and integer $k$, $1\le k \le \frac{n+1}{2}$, we have 
\[\|\Compress_k(w)\| \le \beta_{2k-1}(w).\]
\end{lemma}
\begin{proof}
Let $P$ be the set of positions covered by the leftmost occurrences of subwords from $\Compress_k(w)$.
By construction, $\|\Compress_k(w)\|=|P|$ and $i\in P$ if and only if
$i$ is included in the leftmost occurrence of some subword $s\in \F_{k}(w)$.
If $k\le i \le n-k+1$, then $i$ is the midpoint of a length-$(2k-1)$ fragment $w[i-k+1\dotdot  i+k-1]$,
which covers all length-$k$ fragments of $w$ containing position $i$ including the leftmost occurrence of $s\in \F_k(w)$.
Thus, $w[i-k+1 \dotdot i+k-1]$ is also the leftmost occurrence of the corresponding subword of length $2k-1$.
This yields an injective mapping from the set of positions $i\in P\cap [k\dotdot n-k+1]$ to the family $\F_{2k-1}(w)$.
The remaining positions ($i<k$ and $i>n-k+1$) account for the extra term $2k-2=\beta_{2k-1}(w)-|\F_{2k-1}(w)|$ in the upper bound.
\end{proof}

From the last two lemmas we obtain the following corollary that conveys the intuition behind the main structural theorem in the following section.

\begin{corollary}\label{cor:wr}
  Let $1 \le k \le \frac{n+3}{4}$ and $S=\Compress_{2k-1}(w)$. Then
  \[\Seeds_{\le k}(w) = \bigcap_{s \in S} \Seeds_{\le k}(s) \quad\text{and}\quad \|S\| \le \beta_{4k-3}(w).\]
\end{corollary}

\section{Main Algorithm}\label{sec:main-algorithm}
The main algorithm is a recursive procedure based on a structural theorem which combines the results of the previous section.

  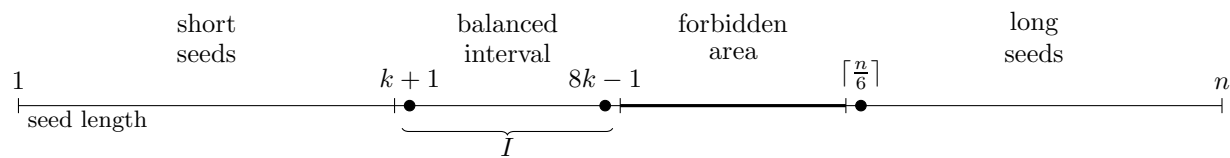
\begin{figure}[ht]
  \begin{center}
    \begin{tikzpicture}
  \draw (-1,0) -- (15,0);
  \draw (-1,-0.1) -- (-1,0.1);
  \draw (4,-0.1) -- (4,0.1);
  \draw (7,-0.1) -- (7,0.1);
  \draw (10,-0.1) -- (10,0.1);
  \draw (15,-0.1) -- (15,0.1);
  \draw (4.2,0) node[above=0.1cm] {$k+1$};
  \draw (6.8,0) node[above=0.1cm] {$8k-1$};
  \filldraw (4.2,0) circle (0.07cm);
  \filldraw (6.8,0) circle (0.07cm);
  \draw (10.2,0) node[above=0.1cm] {$\ceil{\tfrac{n}{6}}$};
  \filldraw (10.2,0) circle (0.07cm);
  \draw (-1,0) node[above=0.1cm] {$1$};
  \draw (15,0) node[above=0.1cm] {$n$};
  \draw[yshift=0.3cm] (8.5,0.8) node {forbidden};
  \draw[yshift=0.3cm] (8.5,0.4) node {area};
  \draw[decorate, decoration=brace] (6.9,-0.3) -- node[below] {$I$} (4.1,-0.3);
  \draw[yshift=0.3cm] (5.5,0.8) node{balanced};
  \draw[yshift=0.3cm] (5.5,0.4) node{interval};
  \draw[very thick] (7,0) -- (10,0);
  \draw[yshift=0.3cm] (12.5,0.8) node{long};
  \draw[yshift=0.3cm] (12.5,0.4) node{seeds};
  \draw[yshift=0.3cm] (1.5,0.8) node {short};
  \draw[yshift=0.3cm] (1.5,0.4) node {seeds};
  \draw (-1,-0.2) node[right] {\small seed length};
  \end{tikzpicture}
\vspace*{-.6cm}
  \end{center}
  \caption{
    Illustration of \cref{thm:IJS}(A) in case that $8k \le \ceil{\frac{n}{6}}$.
    There is no seed of length in the forbidden area.
The computation of all seeds is split into recursive computation of
short seeds (of length at most $k$) in subwords $s\in S$, of seeds with lengths in a balanced interval $I$, and of
long seeds.
    \label{fig:prosta}
  }
  \end{figure}

\begin{theorem}[\bf Decomposition Theorem]\label{thm:IJS}\mbox{ \ }\\
Consider a word $w$ of length $n$. If $n\le 6$ or $|\AAlph(w)|> \frac23 n$, then $\Seeds(w)=\Seeds_{\ge \frac16 n}(w)$. 
Otherwise, we can compute in linear time a balanced interval $I$ and a family $S$ of subwords of $w$ such that
\[\text{\bf (A)}\ \  \Seeds(w)\,=\,\Seeds_{\ge \frac16 n}(w)\cup 
\Seeds_I(w)\, \cup \, \bigcap_{s\in S}\, \Seeds_{\le k}(s)\qquad\text{and}\qquad
\text{\bf (B)}\ \ \|S\| \le \tfrac23 n.\]
\end{theorem}
\begin{proof} 
We take
\begin{align*}
k&=\max\left\{\ell\,:\,1 \le \ell < \ceil{\tfrac{n}{6}}\text{ and } \beta_{4\ell-3}(w)\le \tfrac23 n\right\},\\
S&=\Compress_{2k-1}(w),\\
I&=[k+1 \dotdot \min(8k,\ceil{\tfrac{n}{6}})-1].
\end{align*}

As for the first part of the statement, note that $\Seeds(w)=\Seeds_{\ge \frac16n}(w)$ holds trivially if $n\le 6$.
  On the other hand, if $|\AAlph(w)|=\beta_{1}(w)>\frac23n$, then the equality follows directly from the Gap Lemma (\cref{lem:mks}).

Hence we can assume further that $n>6$ and $\beta_1(w)\le \frac23n$.

\paragraph{Correctness of (B).} 
 Now  we can guarantee that $k$ is well-defined
  and that it satisfies $\beta_{4k-3}(w)\le \frac23 n$. 
  Then Compression Lemma (Lemma~\ref{lem:mkbk}) implies that $S$ satisfies the condition 
{\bf (B)}.

\paragraph{Correctness of (A).} 
To prove condition {\bf (A)}, we observe that the Reduction Lemma (\cref{lem:red}) yields
 $\Seeds_{\le k}(w)=\bigcap_{s\in S} \Seeds_{\le k}(s)$.
 Thus, it is enough to prove that 
$\Seeds_{\ge k+1}(w)= \Seeds_{\ge \frac16 n}(w)\cup \Seeds_I(w)$,
i.e., that there is no seed $v$ with $8k \le |v| < \ceil{\frac{n}{6}}$.
This claim holds trivially for $k=\ceil{\tfrac{n}{6}}-1$. 
Otherwise,  $\beta_{4k+1}>\frac{2}{3}n$, and the claim follows directly from  the Gap Lemma 
(\cref{lem:mks}).

\paragraph{Computing $k$ and $S$.}
Let us start with the following claim.

\begin{claim}\mbox{ \ }\\
  The sequence $|\F_m(w)|$ (consequently, also the sequence $\beta_m(w)$) can be computed in linear time.
\end{claim}
\begin{proof}
  Each equivalence class $E=\pack(i, j_1,j_2)$ contributes a single subword in $\F_m(w)$ for each $m \in [j_1-i+1 \dotdot j_2-i+1]$.
  We obtain at most $2n$ such intervals; $|\F_m(w)|$ is then the number of intervals that contain the element $m$.
  This quantity can be computed in $\Oh(n)$ time using an auxiliary array $A$ of size~$n$ (initially set to zeroes).
  For an interval $[a\dotdot b]$, $A[a]$ is incremented and $A[b+1]$ is decremented.
  Then $|\F_m(w)|=A[1]+\dots+A[m]$.
\end{proof}
To retrieve $S=\Compress_{2k-1}(w)$, we iterate again over the equivalence classes $E=\pack(i,j_1,j_2)$.
If $2k-1 \in [j_1-i+1 \dotdot j_2-i+1]$, we mark the position $\min P$, where $P=\Occ(v)$ for $v\in E$.
Next, we scan the text marking the first $2k-2$ positions in each gap between subsequent already marked positions
and the first $2k-2$ positions after the final already marked position.
After these two phases, we build a word $s\in S$ out of each maximal region of marked positions.
\end{proof}

The algorithm computing seeds relies on \cref{thm:IJS},
with the \LSeedsRep procedure applied to compute $\Seeds_{\ge \frac16 n}(n)$,
and recursive calls made to determine $\Seeds(s)$ for each $s\in S$.
Finally, a package representation of $\Seeds_I(w)$ is computed using \BSeedsRep.
  
    \SetKw{KwReturn}{return}
    \RestyleAlgo{ruled}
    \SetSideCommentRight
    \begin{algorithm*}[h]\label{alg:quasi}
    \caption{Recursive procedure $\FSeeds(w)$}
    \DontPrintSemicolon
      \vskip 0.1cm
    \KwIn{A word $w$ of length $n$.}
    \KwOut{An $\Oh(n)$-size package representation of $\Seeds(w)$.
    }\vskip 0.2cm
    \lIf{$n \le 6$ \KwSty{or} $|\AAlph(w)|>\frac23 n$}{\KwReturn $\LSeedsRep(\ceil{\frac{n}{6}}, w)$}
    \vskip 0.2cm
      Let $I$, $S$ 
      be as in the proof of the Decomposition Theorem\;
      \vskip 0.1cm
       \ForEach{$v\in S$}
        {\vskip 0.1cm
          $R_v:=\FSeeds(v)$}
        \vskip 0.2cm
      $R \;:=\;\Combine(\,\{\,R_v\,:\, v \in S\,\}\,)$\;
      \vskip 0.1cm
      Remove from $R$ seeds of length at least $\min I$\;
      \vskip 0.1cm
    \KwReturn $R\cup \LSeedsRep(\ceil{\frac{n}{6}}, w) 
    \cup \BSeedsRep(I,w)$ \;\vskip 0.1cm
    \hspace*{1cm} (The three package representations contain distinct seeds)
    \end{algorithm*}

\begin{theorem}[\bf Main Result]\label{thm:main}
  An $\Oh(n)$-size package representation of the set $\Seeds(w)$ of all seeds of a given length-$n$ word $w$ can be found in $\Oh(n)$ time.
  In particular, a shortest seed and the total number of seeds 
can be computed within the same time complexity.
\end{theorem}  
   
    \begin{proof}
      Correctness of the algorithm \FSeeds follows immediately from \cref{thm:IJS}.
      To bound the running time, let us denote by $T(n)$ the maximum number of operations performed by the \FSeeds function executed for
      a word of length $n$.
      Due to \cref{thm:IJS,lem:combine,lem:qseedsrep,lem:long},
      \[T(n) = \Oh(n) + \sum_{i} T(n_i), \quad\mbox{where}\ \sum n_i \le \tfrac23 n.\]
      This recurrence yields $T(n) = \Oh(n)$.
    \end{proof}

   \section{Implementation of the Operation $\Combine$}\label{sec:merge}
    \newcommand{\wanc}{\mathtt{ancestor}}
    \newcommand{\wgt}{\mathsf{weight}}

    We describe here the missing part of the algorithm, which 
is based on some rather technical computations on weighted trees.
First, our algorithm requires an efficient solution to the so-called weighted ancestor queries problem.
    Thus, in \cref{subsec:waq} we present an optimal offline procedure answering weighted ancestor queries in an arbitrary weighted tree
    with polynomially bounded weights.
    
    \subsection{Offline Weighted Ancestor Queries}\label{subsec:waq}
    In the weighted ancestor problem, introduced by Farach and Muthukrishnan~\cite{DBLP:conf/cpm/FarachM96} (see also \cite{DBLP:conf/esa/GawrychowskiLN14}),
    we consider a rooted tree $T$ with an integer weight function $\wgt$ defined on the nodes.
    The weight of the root must be zero and the weight of any other node must be strictly larger than the weight
    of its parent.      
    A classic example is any compacted trie with the weight of a node defined as the length of its value.
    
    The weighed ancestor queries, given a node $\nu$ and an integer value $\ell\le \wgt(\nu)$,
    ask for the highest ancestor $\mu$ of $\nu$ such that $\wgt(\mu)\ge \ell$, i.e.,
    such an ancestor $\mu$ that $\wgt(\mu)\ge \ell$ and $\wgt(\mu)$ is smallest possible.
    We denote the node $\mu$ as $\wanc(\nu,\ell)$.

    Weighted ancestor queries in the online setting can be answered in $\Oh(\log \log n)$ time after $\Oh(n)$-time preprocessing~\cite{DBLP:journals/talg/AmirLLS07}.
    In the special case of the weighted tree being a suffix tree of a word, they can be answered in $\Oh(1)$ time with a data structure
    of $\Oh(n)$ space \cite{DBLP:conf/esa/GawrychowskiLN14}.
    Nevertheless, no efficient construction of this data structure is known.
    Below we show that if all the queries are given offline and the weights are polynomially bounded, then $q$ queries can be answered in $\Oh(n+q)$ time.

    Let us first recall the classic union-find data structure.
    It maintains a partition $\SS$ of $[1\dotdot n]$.
	  Each set $S\in \SS$ has an identifier $\id(S)\in S$.
	  Initially, $\SS$ is a partition into singletons and $\id(\{i\})=i$ for $1\le i \le n$.
	  
    The union-find data structure supports $\findop(i)$ queries which, given an integer $i\in [1 \dotdot n]$,
	  return the identifier $\id(S)$ of the set $S\in \SS$ containing $i$.
	  Moreover, a $\unionop(i_1,i_2)$ operation, given integers $i_1,i_2\in  [1 \dotdot n]$,
	  replaces the sets $S_1,S_2\in \SS$ such that $i_1\in S_1$ and $i_2\in S_2$ with their union $S_1\cup S_2$.
	  Note that $\unionop(i_1,i_2)$ is void if $S_1=S_2$, i.e., if $\findop(i_1)=\findop(i_2)$.
	  
	  We will only encounter \emph{linear} union-find instances,
	  in which sets $S\in \SS$ are formed by consecutive integers and $\id(S)=\min(S)$.
	  In other words, $\unionop(i_1,i_2)$ is allowed for $i_1<i_2$ only if $\findop(i_1)=\findop(\findop(i_2)-1)$.
    For such instances, the union-find operations can be implemented in amortized $\Oh(1)$ time.
	  
	  \begin{lemma}[Gabow and Tarjan~\cite{UnionFind1985}]\label{lem:uf}
	  A sequence of $m$ linear union-find operations on a partition of $ [1 \dotdot n]$
	  can be implemented in $\Oh(n+m)$ time.	  
	  \end{lemma}
	  
	  We are now ready to describe an efficient offline procedure answering weighted ancestor queries.
    
    \begin{lemma}\label{lem:wla}
    Given a collection $Q$ of weighted ancestor queries on a weighted tree $T$ on $n$ nodes with integer weights up to $(n+|Q|)^{\Oh(1)}$,
    all the queries from $Q$ can be answered in $\Oh(n+|Q|)$ time.
    \end{lemma}
    \begin{proof}
	  We process the tree and the queries according to non-increasing weights.
	  We maintain a union-find data structure which stores a partition of the set $V(T)$
	  of nodes of $T$. After the nodes with weight $\ell$ have been processed, each partition class is either a singleton of a node
	  $\mu$ such that $\wgt(\mu)<\ell$, or consists of the nodes of a subtree rooted at a node $\mu$ such that $\wgt(\mu)\ge \ell$.
	  In either case, $\mu$ is the identifier of the set.
	  
	  Note that, in order to update the partition for the next smaller value of $\ell$,
	  for each node $\mu$ with weight $\ell$ it suffices to union the singleton $\{\mu\}$
	  with the node sets of subtrees rooted at the children of $\mu$.
	  Moreover, observe that after processing nodes at level $\ell$,
	  for each node $\nu$, its ancestors $\mu$ with $\wgt(\mu)<\ell$
	  form singletons, while ancestors $\mu$ with $\wgt(\mu)\ge \ell$ belong to the same class as $\nu$.
	  Hence, the identifier of this class is the highest ancestor of $\mu$ with $\wgt(\mu)\ge \ell$,
	  i.e., $\wanc(\nu,\ell)=\findop(\nu)$.
	  Consequently, the weighted ancestor queries can be answered using \cref{alg:owa}.

    \begin{algorithm}[ht]
    \vskip 0.1cm
    $W_T = \{\wgt(\mu): \mu\in V(T)\}$\;
    $W_Q = \{\ell : (\nu,\ell)\in Q\}$\;\vskip 0.1cm
    \ForEach{$\ell\in W_T\cup W_Q$ in the decreasing order}{
    	\ForEach{$\mu\in V(T) : \wgt(\mu)=\ell$}{
    		\ForEach{$\nu: \text{child of }\mu$ in the left-to-right order}{
    			$\unionop(\mu,\nu)$\;
    		}\vskip 0.1cm
    		
    	}
    	\ForEach{$\nu : (\nu,\ell)\in Q$}{
    		Report $\wanc(\nu,\ell):=\findop(\nu)$\;
    	}
    }
    \caption{Offline weighted ancestors for a weighted tree $T$ and a set of queries $Q$}\label{alg:owa}
    \end{algorithm}
    
    Next, we shall prove that \cref{alg:owa} can be implemented in $\Oh(n+|Q|)$ time.
    Since node weights and query weights are integers bounded by $(n+|Q|)^{\Oh(1)}$, they can be sorted using radix sort in $\Oh(n+|Q|)$ time.
    For union-find operations, we need to identify nodes with integers $ [1 \dotdot n]$.
    We use the pre-order identifiers as they guarantee that each partition class 
    (which can be either a singleton or the node set of a subtree rooted at a given node) consists of consecutive integers.
    With $n-1$ $\unionop$ operations and $|Q|$ $\findop$ operations, \cref{lem:uf} guarantees $\Oh(n+|Q|)$ overall running time of the union-find data structure. 
    \end{proof}

    We apply offline weighted ancestor queries to the suffix tree to obtain the following algorithmic tool.

    \begin{corollary}\label{cor:wla}
    Given a collection of subwords $s_1,\ldots,s_k$ of a word $w$ of length $n$, each represented by an occurrence in $w$,
    in $\Oh(n+k)$ total time we can compute the locus of each subword $s_i$ in the suffix tree of $w$.
    Moreover, these loci can be made explicit in $\Oh(n+k)$ extra time.
    \end{corollary}
    \begin{proof}
    Let $T$ be the suffix tree of $w$.
    Assume that $s_i = w[a\dotdot b]$ and consider the following nodes: the node $\mu$ representing $w[a\dotdot n]$ (the terminal node of $T$ annotated with $a$)
    and $\nu=\wanc(\mu,b-a+1)$.
    If we denote by $d$ the depth of $\nu$, then the locus of $s_i$ is $(\nu,d-(b-a+1))$.
    By \cref{lem:wla}, the loci of $s_i$ can be computed in $\Oh(n+k)$ time.

    In order to make the corresponding implicit nodes explicit, we need to group them according to the nearest explicit descendant
    and sort them by distances from that node.
    This can be implemented in $\Oh(n+k)$ time via radix sort.
    Then we simply create the explicit nodes in the obtained order.
    \end{proof}

\subsection{Intersection of Families of Paths in a Tree}
    Let us introduce one more abstract operation on a rooted tree $T$.
    A \emph{path family} is a family of pairwise disjoint paths in $T$, each leading downwards.
    A path family is called \emph{minimal} if there is no other path family covering the same
    set of nodes in $T$ and consisting of a smaller number of paths.

  \begin{figure}[htpb]
  \begin{center}
      \begin{tikzpicture}[scale=1.2]

    \foreach \x in {0,4,9}{
      \begin{scope}[xshift=\x cm]
        \draw (0,0) -- (1,-1) -- (1.5,-2)  (1,-1) -- (0.5,-2)  (0,0) -- (-1,-1) -- (-1.5,-2)  (-1,-1) -- (-0.5,-2);
        \foreach \x/\y in {
          0/0,
            -0.5/-0.5, -1/-1, 0.5/-0.5, 1/-1,
            -1.5/-2, -0.5/-2, 0.5/-2, 1.5/-2, -1.25/-1.5, -0.75/-1.5, 0.75/-1.5, 1.25/-1.5
        }{
          \filldraw[white] (\x,\y) circle (0.07cm);
          \draw (\x,\y) circle (0.07cm);
        }
      \end{scope}
    }
    \draw[very thick,xshift=-0.2cm,yshift=0.1cm,rounded corners=20pt] (-0.5,-0.5) -- (-1,-1) -- (-1.25,-1.5);
    \draw[very thick,xshift=0.2cm,yshift=0.1cm,rounded corners=20pt] (0.5,-0.5) -- (1,-1) -- (1.5,-2);
    \draw[very thick,xshift=0.2cm,yshift=0.1cm] (-0.5,-2) -- (-0.75,-1.5);

    \begin{scope}[xshift=4cm]
    \draw[very thick,xshift=-0.1cm,yshift=0.1cm,rounded corners=20pt] (0.5,-0.7) -- (1,-1) -- (0.75,-1.5);
    \draw[very thick,xshift=0.1cm,yshift=0.1cm,rounded corners=20pt] (-0.5,-0.7) -- (-1,-1) -- (-0.5,-2);
    \draw[very thick,xshift=0.2cm,yshift=0.1cm] (1,-1) -- (1.5,-2);
    \end{scope}

    \draw[very thick,-latex,xshift=-1cm] (6.7,-1) -- node[above] {$\Paths$} (8.3,-1);
  
    \begin{scope}[xshift=9cm]
    \draw[very thick,xshift=-0.2cm,yshift=0.1cm] (-0.5,-0.5) -- (-1,-1);
    \draw[very thick,xshift=0.2cm,yshift=0.1cm] (-0.75,-1.5) -- (-0.5,-2);
    \draw[very thick,xshift=0.2cm,yshift=0.1cm,rounded corners=20pt] (0.5,-0.5) -- (1,-1) -- (1.5,-2);
    \end{scope}

  \end{tikzpicture}
  \end{center}
  \caption{
    The $\Paths_T(P_1,P_2)$ operation: three copies of the same rooted tree $T$,
    the first two show $P_1$ and $P_2$, and the third one shows $\Paths_T(P_1,P_2)$.
    \label{fig:paths}
  }
  \end{figure}
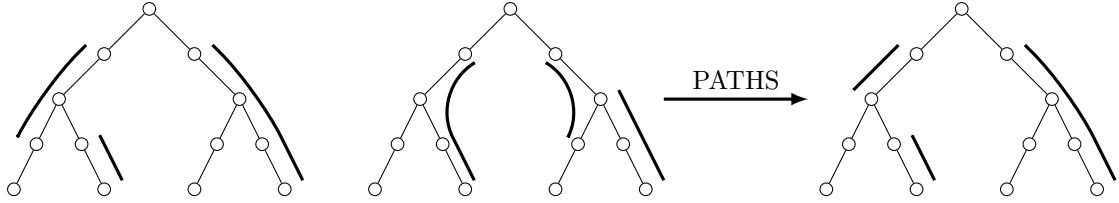
    Let $P_1,\ldots,P_m$ be path families in $T$.
    Then by $\Paths_T(P_1,\ldots,P_m)$ we denote a minimal path family representing the nodes
    covered by all the families $P_1,\ldots,P_m$; see \cref{fig:paths}.

    \begin{lemma}\label{lem:paths}
      The family $\Paths_T(P_1,\ldots,P_m)$ can be computed in linear time with respect to the size of the tree $T$,
      the number $m$, and total number of paths in all families $P_i$.
    \end{lemma}
    \begin{proof}
      For each node $\mu$ in $T$, we would like to compute a value, denoted $S[\mu]$, that is equal to the number of paths in all families $P_i$ that contain node $\mu$.
      Observe that a node $\mu$ is covered by all the families $P_i$ if and only if $S[\mu]=m$.
      
      For each node $\nu$ in $T$, we will store a counter $C[\nu]$ so that, for a node $\mu$,
      $S[\mu]$ is equal to the sum of values $C[\nu]$ across the nodes $\nu$ in the subtree of $\mu$.
      The counters $C$ are initially set to zeroes.
      For each path leading from $\mu$ down to $\nu$ in $P_i$, we increment the counter $C$ at $\nu$ and decrement the counter at the parent of $\mu$
      (unless $\mu$ is the root).
      Next, for each node $\mu$, we compute $S[\mu]$ as the sum of values $C[\nu]$ across the nodes $\nu$ in the subtree of $\mu$.
      This is done in a bottom-up fashion using the $S[\cdot]$ values of the children of $\mu$.
      
      This way, we compute all the nodes represented by $\Paths_T(P_1,\ldots,P_m)$.
      Finally, in order to find the minimal path family covering all these nodes, we repeat the following process traversing the tree in a bottom-up order:
      If the value $S[\mu]$ is $m$, we create a single-node path $\{\mu\}$.
      If also $S[\nu]=m$ for a child $\nu$ of $\mu$, we merge the paths containing these two nodes.
      (We choose the child arbitrarily if $S[\nu]=m$ for several children.)
    \end{proof}
 
    \subsection{Proof of \cref{lem:combine}}\label{subsec:combine}
    We are now ready to provide an efficient implementation of the $\Combine$ operation.
    \lemcombine*
    \begin{proof}
    We reduce the problem to computing $\Paths_T(P_1,\ldots,P_m)$ for some path families $P_1,\ldots,P_m$.

    We first apply \cref{cor:wla} for subwords $w[i\dotdot j_1]$ and $w[i\dotdot j_2]$ across all packages $\pack(i,j_1,j_2)$ in $R_1,\ldots,R_m$,
    extending the set of explicit nodes of the suffix tree $T$ of $w$ by the obtained loci.
    For each package, we create a path connecting the corresponding two loci.
    The packages in a package representation are disjoint, so for each $R_i$ this process results in a path family.
    We apply \cref{lem:paths} to compute a minimal path family $P=\Paths_T(P_1,\ldots,P_m)$.

    Finally, for each path in $P$, we create a package in $T$.
    We assume that each node stores the label $\ell$ of any terminal node in its subtree.
    If a path in $P$ connects two nodes at depths $d_1 \le d_2$, and the second one stores the label $\ell$,
    then we create a package $\pack(\ell,\ell+d_1-1,\ell+d_2-1)$.
    \end{proof}

    \paragraph{Acknowledgements}
    The authors thank Patryk Czajka for suggesting a simplification of the algorithm in \cref{subsec:lc} (\cref{lem:lcand}).

    Jakub Radoszewski was supported by the ``Algorithms for text processing with errors and uncertainties'' project carried out
    within the HOMING program of the Foundation for Polish Science co-financed by the European Union under the European Regional Development Fund.

\bibliographystyle{plainurl}
\bibliography{seeds}

\begin{thebibliography}{10}

\bibitem{DBLP:journals/talg/AmirLLS07}
Amihood Amir, Gad~M. Landau, Moshe Lewenstein, and Dina Sokol.
\newblock Dynamic text and static pattern matching.
\newblock {\em {ACM} Trans. Algorithms}, 3(2):19, 2007.
\newblock \href {http://dx.doi.org/10.1145/1240233.1240242}
  {\path{doi:10.1145/1240233.1240242}}.

\bibitem{DBLP:conf/cpm/AmirLLLP17}
Amihood Amir, Avivit Levy, Moshe Lewenstein, Ronit Lubin, and Benny Porat.
\newblock Can we recover the cover?
\newblock In Juha K{\"{a}}rkk{\"{a}}inen, Jakub Radoszewski, and Wojciech
  Rytter, editors, {\em 28th Annual Symposium on Combinatorial Pattern
  Matching, {CPM} 2017}, volume~78 of {\em LIPIcs}, pages 25:1--25:15. Schloss
  Dagstuhl - Leibniz-Zentrum fuer Informatik, 2017.
\newblock \href {http://dx.doi.org/10.4230/LIPIcs.CPM.2017.25}
  {\path{doi:10.4230/LIPIcs.CPM.2017.25}}.

\bibitem{DBLP:conf/cpm/AmirLLP17}
Amihood Amir, Avivit Levy, Ronit Lubin, and Ely Porat.
\newblock Approximate cover of strings.
\newblock In Juha K{\"{a}}rkk{\"{a}}inen, Jakub Radoszewski, and Wojciech
  Rytter, editors, {\em 28th Annual Symposium on Combinatorial Pattern
  Matching, {CPM} 2017}, volume~78 of {\em LIPIcs}, pages 26:1--26:14. Schloss
  Dagstuhl - Leibniz-Zentrum fuer Informatik, 2017.
\newblock \href {http://dx.doi.org/10.4230/LIPIcs.CPM.2017.26}
  {\path{doi:10.4230/LIPIcs.CPM.2017.26}}.

\bibitem{DBLP:conf/cpm/AmirLP18}
Amihood Amir, Avivit Levy, and Ely Porat.
\newblock Quasi-periodicity under mismatch errors.
\newblock In Gonzalo Navarro, David Sankoff, and Binhai Zhu, editors, {\em
  Annual Symposium on Combinatorial Pattern Matching, {CPM} 2018}, volume 105
  of {\em LIPIcs}, pages 4:1--4:15. Schloss Dagstuhl - Leibniz-Zentrum fuer
  Informatik, 2018.
\newblock \href {http://dx.doi.org/10.4230/LIPIcs.CPM.2018.4}
  {\path{doi:10.4230/LIPIcs.CPM.2018.4}}.

\bibitem{DBLP:journals/tcs/ApostolicoE93}
Alberto Apostolico and Andrzej Ehrenfeucht.
\newblock Efficient detection of quasiperiodicities in strings.
\newblock {\em Theoretical Computer Science}, 119(2):247--265, 1993.
\newblock \href {http://dx.doi.org/10.1016/0304-3975(93)90159-Q}
  {\path{doi:10.1016/0304-3975(93)90159-Q}}.

\bibitem{DBLP:journals/ipl/ApostolicoFI91}
Alberto Apostolico, Martin Farach, and Costas~S. Iliopoulos.
\newblock Optimal superprimitivity testing for strings.
\newblock {\em Information Processing Letters}, 39(1):17--20, 1991.
\newblock \href {http://dx.doi.org/10.1016/0020-0190(91)90056-N}
  {\path{doi:10.1016/0020-0190(91)90056-N}}.

\bibitem{DBLP:journals/iandc/BerkmanIP95}
Omer Berkman, Costas~S. Iliopoulos, and Kunsoo Park.
\newblock The subtree max gap problem with application to parallel string
  covering.
\newblock {\em Information and Computation}, 123(1):127--137, 1995.
\newblock \href {http://dx.doi.org/10.1006/inco.1995.1162}
  {\path{doi:10.1006/inco.1995.1162}}.

\bibitem{DBLP:journals/ipl/Breslauer92}
Dany Breslauer.
\newblock An on-line string superprimitivity test.
\newblock {\em Information Processing Letters}, 44(6):345--347, 1992.
\newblock \href {http://dx.doi.org/10.1016/0020-0190(92)90111-8}
  {\path{doi:10.1016/0020-0190(92)90111-8}}.

\bibitem{DBLP:conf/cpm/BrodalP00}
Gerth~St{\o}lting Brodal and Christian N.~S. Pedersen.
\newblock Finding maximal quasiperiodicities in strings.
\newblock In Raffaele Giancarlo and David Sankoff, editors, {\em Combinatorial
  Pattern Matching, {CPM} 2000}, volume 1848 of {\em Lecture Notes in Computer
  Science}, pages 397--411. Springer, 2000.
\newblock \href {http://dx.doi.org/10.1007/3-540-45123-4_33}
  {\path{doi:10.1007/3-540-45123-4_33}}.

\bibitem{DBLP:journals/jalc/ChristodoulakisIPS05}
Manolis Christodoulakis, Costas~S. Iliopoulos, Kunsoo Park, and Jeong~Seop Sim.
\newblock Approximate seeds of strings.
\newblock {\em Journal of Automata, Languages and Combinatorics},
  10(5/6):609--626, 2005.

\bibitem{DBLP:journals/tcs/ChristouCIKPRRSW13}
Michalis Christou, Maxime Crochemore, Costas~S. Iliopoulos, Marcin Kubica,
  Solon~P. Pissis, Jakub Radoszewski, Wojciech Rytter, Bartosz Szreder, and
  Tomasz Waleń.
\newblock Efficient seed computation revisited.
\newblock {\em Theoretical Computer Science}, 483:171--181, 2013.
\newblock \href {http://dx.doi.org/10.1016/j.tcs.2011.12.078}
  {\path{doi:10.1016/j.tcs.2011.12.078}}.

\bibitem{DBLP:journals/jalc/ColeIMSY05}
Richard Cole, Costas~S. Iliopoulos, Manal Mohamed, William~F. Smyth, and
  Lu~Yang.
\newblock The complexity of the minimum k-cover problem.
\newblock {\em Journal of Automata, Languages and Combinatorics},
  10(5/6):641--653, 2005.

\bibitem{AlgorithmsOnStrings}
Maxime Crochemore, Christophe Hancart, and Thierry Lecroq.
\newblock {\em Algorithms on strings}.
\newblock Cambridge University Press, 2007.
\newblock \href {http://dx.doi.org/10.1017/cbo9780511546853}
  {\path{doi:10.1017/cbo9780511546853}}.

\bibitem{Jewels}
Maxime Crochemore and Wojciech Rytter.
\newblock {\em Jewels of Stringology}.
\newblock World Scientific, 2003.
\newblock \href {http://dx.doi.org/10.1142/4838} {\path{doi:10.1142/4838}}.

\bibitem{DBLP:journals/jacm/Farach-ColtonFM00}
Martin Farach{-}Colton, Paolo Ferragina, and S.~Muthukrishnan.
\newblock On the sorting-complexity of suffix tree construction.
\newblock {\em Journal of the {ACM}}, 47(6):987--1011, 2000.
\newblock \href {http://dx.doi.org/10.1145/355541.355547}
  {\path{doi:10.1145/355541.355547}}.

\bibitem{DBLP:conf/cpm/FarachM96}
Martin Farach-Colton and S.~Muthukrishnan.
\newblock Perfect hashing for strings: {F}ormalization and algorithms.
\newblock In Daniel~S. Hirschberg and Eugene~W. Myers, editors, {\em
  Combinatorial Pattern Matching, {CPM} 1996}, volume 1075 of {\em {Lecture
  Notes in Computer Science}}, pages 130--140. Springer, 1996.
\newblock \href {http://dx.doi.org/10.1007/3-540-61258-0_11}
  {\path{doi:10.1007/3-540-61258-0_11}}.

\bibitem{fine1965uniqueness}
Nathan~J. Fine and Herbert~S. Wilf.
\newblock Uniqueness theorems for periodic functions.
\newblock {\em Proceedings of the American Mathematical Society},
  16(1):109--114, 1965.
\newblock \href {http://dx.doi.org/10.2307/2034009}
  {\path{doi:10.2307/2034009}}.

\bibitem{DBLP:journals/tcs/FlouriIKPPST13}
Tom{\'{a}}s Flouri, Costas~S. Iliopoulos, Tomasz Kociumaka, Solon~P. Pissis,
  Simon~J. Puglisi, William~F. Smyth, and Wojciech Tyczy\'nski.
\newblock Enhanced string covering.
\newblock {\em Theoretical Computer Science}, 506:102--114, 2013.
\newblock \href {http://dx.doi.org/10.1016/j.tcs.2013.08.013}
  {\path{doi:10.1016/j.tcs.2013.08.013}}.

\bibitem{DBLP:conf/stoc/GabowBT84}
Harold~N. Gabow, Jon~Louis Bentley, and Robert~Endre Tarjan.
\newblock Scaling and related techniques for geometry problems.
\newblock In Richard~A. DeMillo, editor, {\em 16th Annual {ACM} Symposium on
  Theory of Computing, {STOC} 1984}, pages 135--143. {ACM}, 1984.
\newblock \href {http://dx.doi.org/10.1145/800057.808675}
  {\path{doi:10.1145/800057.808675}}.

\bibitem{UnionFind1985}
Harold~N. Gabow and Roberd~E. Tarjan.
\newblock A linear-time algorithm for a special case of disjoint set union.
\newblock {\em Journal of Computer and System Sciences}, 30(2):209--221, 1985.
\newblock \href {http://dx.doi.org/10.1016/0022-0000(85)90014-5}
  {\path{doi:10.1016/0022-0000(85)90014-5}}.

\bibitem{DBLP:conf/esa/GawrychowskiLN14}
Paweł Gawrychowski, Moshe Lewenstein, and Patrick~K. Nicholson.
\newblock Weighted ancestors in suffix trees.
\newblock In Andreas~S. Schulz and Dorothea Wagner, editors, {\em Algorithms,
  {ESA} 2014}, volume 8737 of {\em {Lecture Notes in Computer Science}}, pages
  455--466. Springer, 2014.
\newblock \href {http://dx.doi.org/10.1007/978-3-662-44777-2_38}
  {\path{doi:10.1007/978-3-662-44777-2_38}}.

\bibitem{DBLP:conf/aaim/GuoZI06}
Qing Guo, Hui Zhang, and Costas~S. Iliopoulos.
\newblock Computing the {$\lambda$}-seeds of a string.
\newblock In Siu{-}Wing Cheng and Chung~Keung Poon, editors, {\em Algorithmic
  Aspects in Information and Management, {AAIM} 2006}, volume 4041 of {\em
  Lecture Notes in Computer Science}, pages 303--313. Springer, 2006.
\newblock \href {http://dx.doi.org/10.1007/11775096_28}
  {\path{doi:10.1007/11775096_28}}.

\bibitem{DBLP:conf/spire/GuoZI06}
Qing Guo, Hui Zhang, and Costas~S. Iliopoulos.
\newblock Computing the minimum approximate $\lambda$-cover of a string.
\newblock In Fabio Crestani, Paolo Ferragina, and Mark Sanderson, editors, {\em
  String Processing and Information Retrieval, 13th International Conference,
  {SPIRE} 2006}, volume 4209 of {\em Lecture Notes in Computer Science}, pages
  49--60. Springer, 2006.
\newblock \href {http://dx.doi.org/10.1007/11880561\_5}
  {\path{doi:10.1007/11880561\_5}}.

\bibitem{DBLP:journals/isci/GuoZI07}
Qing Guo, Hui Zhang, and Costas~S. Iliopoulos.
\newblock Computing the {$\lambda$}-covers of a string.
\newblock {\em Information Sciences}, 177(19):3957--3967, 2007.
\newblock \href {http://dx.doi.org/10.1016/j.ins.2007.02.020}
  {\path{doi:10.1016/j.ins.2007.02.020}}.

\bibitem{DBLP:journals/isci/IliopoulosMS11}
Costas~S. Iliopoulos, Manal Mohamed, and William~F. Smyth.
\newblock New complexity results for the k-covers problem.
\newblock {\em Information Sciences}, 181(12):2571--2575, 2011.
\newblock \href {http://dx.doi.org/10.1016/j.ins.2011.02.009}
  {\path{doi:10.1016/j.ins.2011.02.009}}.

\bibitem{DBLP:journals/algorithmica/IliopoulosMP96}
Costas~S. Iliopoulos, Dennis W.~G. Moore, and Kunsoo Park.
\newblock Covering a string.
\newblock {\em Algorithmica}, 16(3):288--297, 1996.
\newblock \href {http://dx.doi.org/10.1007/BF01955677}
  {\path{doi:10.1007/BF01955677}}.

\bibitem{DBLP:journals/jalc/IliopoulosM99}
Costas~S. Iliopoulos and Laurent Mouchard.
\newblock Quasiperiodicity: From detection to normal forms.
\newblock {\em Journal of Automata, Languages and Combinatorics},
  4(3):213--228, 1999.

\bibitem{IliopoulosSmythKCovers}
Costas~S. Iliopoulos and William~F. Smyth.
\newblock An on-line algorithm of computing a minimum set of k-covers of a
  string.
\newblock In {\em Australasian Workshop on Combinatorial Algorithms, AWOCA
  1998}, pages 97--106, 1998.

\bibitem{DBLP:journals/jacm/KarkkainenSB06}
Juha K{\"{a}}rkk{\"{a}}inen, Peter Sanders, and Stefan Burkhardt.
\newblock Linear work suffix array construction.
\newblock {\em Journal of the {ACM}}, 53(6):918--936, 2006.
\newblock \href {http://dx.doi.org/10.1145/1217856.1217858}
  {\path{doi:10.1145/1217856.1217858}}.

\bibitem{DBLP:conf/soda/KociumakaKRRW12}
Tomasz Kociumaka, Marcin Kubica, Jakub Radoszewski, Wojciech Rytter, and Tomasz
  Waleń.
\newblock A linear time algorithm for seeds computation.
\newblock In Yuval Rabani, editor, {\em 23rd Annual {ACM-SIAM} Symposium on
  Discrete Algorithms, {SODA} 2012}, pages 1095--1112. {SIAM}, 2012.
\newblock \href {http://dx.doi.org/10.1137/1.9781611973099}
  {\path{doi:10.1137/1.9781611973099}}.

\bibitem{DBLP:journals/tcs/KociumakaPRRW16}
Tomasz Kociumaka, Solon~P. Pissis, Jakub Radoszewski, Wojciech Rytter, and
  Tomasz Wale\'n.
\newblock Efficient algorithms for shortest partial seeds in words.
\newblock {\em Theoretical Computer Science}, 710:139--147, 2018.
\newblock \href {http://dx.doi.org/10.1016/j.tcs.2016.11.035}
  {\path{doi:10.1016/j.tcs.2016.11.035}}.

\bibitem{DBLP:journals/algorithmica/KociumakaPRRW15}
Tomasz Kociumaka, Solon~P. Pissis, Jakub Radoszewski, Wojciech Rytter, and
  Tomasz Waleń.
\newblock Fast algorithm for partial covers in words.
\newblock {\em Algorithmica}, 73(1):217--233, 2015.
\newblock \href {http://dx.doi.org/10.1007/s00453-014-9915-3}
  {\path{doi:10.1007/s00453-014-9915-3}}.

\bibitem{DBLP:conf/focs/KolpakovK99}
Roman~M. Kolpakov and Gregory Kucherov.
\newblock Finding maximal repetitions in a word in linear time.
\newblock In {\em 40th Annual Symposium on Foundations of Computer Science,
  {FOCS} 1999}, pages 596--604. {IEEE} Computer Society, 1999.
\newblock \href {http://dx.doi.org/10.1109/SFFCS.1999.814634}
  {\path{doi:10.1109/SFFCS.1999.814634}}.

\bibitem{DBLP:journals/algorithmica/LiS02}
Yin Li and William~F. Smyth.
\newblock Computing the cover array in linear time.
\newblock {\em Algorithmica}, 32(1):95--106, 2002.
\newblock \href {http://dx.doi.org/10.1007/s00453-001-0062-2}
  {\path{doi:10.1007/s00453-001-0062-2}}.

\bibitem{Lothaire2005}
M.~Lothaire.
\newblock {\em Applied Combinatorics on Words}.
\newblock Cambridge University Press, 2005.
\newblock \href {http://dx.doi.org/10.1017/cbo9781107341005}
  {\path{doi:10.1017/cbo9781107341005}}.

\bibitem{DBLP:journals/siamcomp/ManberM93}
Udi Manber and Eugene~W. Myers.
\newblock Suffix arrays: {A} new method for on-line string searches.
\newblock {\em {SIAM} Journal on Computing}, 22(5):935--948, 1993.
\newblock \href {http://dx.doi.org/10.1137/0222058}
  {\path{doi:10.1137/0222058}}.

\bibitem{DBLP:journals/ipl/MooreS94}
Dennis W.~G. Moore and William~F. Smyth.
\newblock An optimal algorithm to compute all the covers of a string.
\newblock {\em Information Processing Letters}, 50(5):239--246, 1994.
\newblock \href {http://dx.doi.org/10.1016/0020-0190(94)00045-X}
  {\path{doi:10.1016/0020-0190(94)00045-X}}.

\bibitem{DBLP:journals/ipl/MooreS95}
Dennis W.~G. Moore and William~F. Smyth.
\newblock A correction to "{A}n optimal algorithm to compute all the covers of
  a string".
\newblock {\em Information Processing Letters}, 54(2):101--103, 1995.
\newblock \href {http://dx.doi.org/10.1016/0020-0190(94)00235-Q}
  {\path{doi:10.1016/0020-0190(94)00235-Q}}.

\bibitem{morris1970linear}
James~H. Morris, Jr. and Vaughan~R. Pratt.
\newblock A linear pattern-matching algorithm.
\newblock Technical Report~40, Department of Computer Science, University of
  California, Berkeley, 1970.

\bibitem{SimParkKimLee}
Jeong~Seop Sim, Kunsoo Park, Sung-Ryul Kim, and Jee-Soo Lee.
\newblock Finding approximate covers of strings.
\newblock {\em Journal of Korea Information Science Society}, 29(1):16--21,
  2002.
\newblock URL:
  \url{http://www.koreascience.or.kr/article/ArticleFullRecord.jsp?cn=JBGHG6_2002_v29n1_16}.

\bibitem{DBLP:journals/tcs/Smyth00}
William~F. Smyth.
\newblock Repetitive perhaps, but certainly not boring.
\newblock {\em Theoretical Computer Science}, 249(2):343--355, 2000.
\newblock \href {http://dx.doi.org/10.1016/S0304-3975(00)00067-0}
  {\path{doi:10.1016/S0304-3975(00)00067-0}}.

\bibitem{DBLP:conf/focs/Weiner73}
Peter Weiner.
\newblock Linear pattern matching algorithms.
\newblock In {\em 14th Annual Symposium on Switching and Automata Theory,
  {SWAT} 1973}, pages 1--11, Washington, DC, USA, 1973. IEEE Computer Society.
\newblock \href {http://dx.doi.org/10.1109/SWAT.1973.13}
  {\path{doi:10.1109/SWAT.1973.13}}.

\end{thebibliography}

\end{document}